\documentclass[11pt]{article}
\usepackage[margin=1in]{geometry}
\usepackage{url, dsfont, dsfont}
\usepackage[round]{natbib}
\usepackage[colorlinks, bookmarksopen, bookmarksnumbered, citecolor=red, urlcolor=red]{hyperref}
\usepackage{verbatim}
\usepackage{xr}
\usepackage{xr-hyper}
\usepackage{xcolor}
\hypersetup{colorlinks=true, linkcolor={red}, citecolor={blue}, urlcolor={red}}
\usepackage{soul}
\AtBeginDocument{\hypersetup{pdfborder={0 0 0}}}
\usepackage{moreverb, subcaption, threeparttable, multicol, multirow, lscape, bm, soul, amsthm, amsfonts, amsmath}
\usepackage[title]{appendix}
\usepackage{mathtools, color, xcolor}
\usepackage{tabularx, ltablex, floatrow, inputenc}
\usepackage[]{graphicx}
\usepackage{color, threeparttable, multirow, multicol, setspace}
\usepackage{bbm, caption, tikz, colortbl, dcolumn, float, enumerate, enumitem, booktabs, makecell}

\newcommand{\bd}{\boldsymbol}
\newcommand{\Ex}{\mathbb{E}}

\newcommand{\Px}{\mathbb{P}}

\newcommand{\mc}{\mathcal}
\newcommand{\mb}{\mathbf}
\newcommand{\WR}{\theta^\text{WR}}
\newcommand{\WRhat}{\widehat\theta^\text{WR}}

\newcommand{\bigCI}{\mathrel{\text{\scalebox{1}{$\perp\mkern-10mu\perp$}}}}

\theoremstyle{thmstyleone}%

\newtheorem{assumption}{Assumption}
\newtheorem{remark}{Remark}

\providecommand{\keywords}[1]{\small\textbf{{Keywords}} #1}

\title{\bf\Large Estimation and Inference of the Win Ratio for Two Hierarchical Endpoints Subject to Censoring and Missing Data  }

\title{\bf Estimation and Inference of the Win Ratio for Two Hierarchical Endpoints Subject to Censoring and Missing Data}
\author{Yi Liu$^{*1,2}$, Huiman Barnhart$^{2,3}$, Sean O'Brien$^{2,3}$, \\ Yuliya Lokhnygina$^{2,3}$, and Roland A. Matsouaka$^{2,3}$ 
\bigskip
\\
\normalsize 
$^{1}$Department of Statistics, North Carolina State University, Raleigh, NC, USA \\
\normalsize 
$^{2}$Duke Clinical Research Institute, Durham, NC, USA \\
\normalsize 
$^{3}$Department of Biostatistics and Bioinformatics, Duke University, Durham, NC, USA 
\bigskip
\\
\normalsize 
$^*$Corresponding author: Yi Liu, \texttt{yliu297@ncsu.edu}
}
 
\date{}

\begin{document}

\maketitle

\begin{abstract}

The win ratio (WR) is a widely used metric to compare treatments in randomized clinical trials with hierarchically ordered endpoints. Counting-based approaches, such as Pocock's algorithm, are the standard for WR estimation. However, this algorithm treats participants with censored or missing data inadequately,  which may lead to biased and inefficient estimates, particularly in the presence of heterogeneous censoring or missing data between treatment groups. Although recent extensions have addressed some of these limitations for hierarchical time-to-event endpoints, no existing methods---aside from the computationally intensive multiple imputation approach---can accommodate settings that include non-survival endpoints that are subject to missing data. In this paper, we propose a simple nonparametric maximum likelihood estimator (NPMLE) of WR for two hierarchical endpoints that are subject to censoring and missing data. Our method uses all observed data, avoids strong parametric assumptions, and comes with a closed-form asymptotic variance estimator. We demonstrate its performance using simulation studies and two data examples, based on the HEART-FID and ISCHEMIA trials. The proposed method provides a consistent estimator, improves estimation efficiency, and is robust under non-informative censoring and missing at random (MAR) assumptions, offering a flexible alternative to existing WR estimation methods. A user-friendly R package, \texttt{WinRS}, is available to facilitate implementation. 

\end{abstract} \hspace{10pt}

\doublespacing

\section{Introduction}\label{sec:intro}

Randomized clinical trials often use composite endpoints to comprehensively assess the effect of treatment on a target population. For example, trials for cardiovascular diseases often evaluate endpoints such as death, stroke, myocardial infarction and quality of life. Incorporating multiple endpoints in a hierarchical structure at the design stage of a trial allows the utilization of more information to assess treatment efficacy and safety \citep{barnhart2025trial, finkelstein1999combining}. The win ratio (WR), introduced by \cite{pocock2012win}, is a useful and germane measure to compare two treatment groups with hierarchical endpoints. 


Pocock's algorithm is widely used to estimate the WR; it is similar to the Finkelstein-Schoenfeld method \citep{finkelstein1999combining} and extends the Wilcoxon-Mann-Whitney test paradigm  to multiple outcomes of different types (categorical, continuous, discrete, ...) through the generalized pairwise comparisons (GPC) framework \citep{pocock2012win,Buyse2025GPC}. This method has been extensively applied in real-world studies \citep{abdalla2016win, anker2009ferric, ferreira2020use, krittayaphong2024components, redfors2020win, maron2020initial, monzo2024use}, and user-friendly software packages are readily available for implementation \citep{cui2025wins, dong2023win, Buyse2025GPC}. However, when data include censored or missing values, which is common in longitudinal and time-to-event settings, Pocock's estimator can be biased \citep{Buyse2025GPC}. 

Suppose we want to compare two treatment options (hereafter ``treatment'' vs. ``placebo'') on a list of prioritized endpoints, ranked from highest to lowest, following a pre-specified hierarchy. We can consider pairs of participants, one from each treatment group, and compare their endpoints sequentially. The key idea of Pocock's method is to count the number of wins, losses, and ties among all possible pairwise comparisons of participants and estimate the WR. Starting with the highest-ranked endpoint, we assess for each pair whether the participant in the treatment group fares better (i.e., ``wins'') or worse (i.e., ``loses''). If the pair is a tie or we are unable to compare, then we carry the comparison on the second highest-ranked endpoint. The process continues until there is a winner or we reach the last endpoint {on the hierarchy} without a clear winner, at which point the pair is declared a tie. {After evaluating all pairs,} we count the total number of {wins}, losses, and ties and estimate the WR as the ratio of the number of wins over the number of losses \citep{pocock2012win}. {Pairs that remain unresolved after exhausting the hierarchy are recorded as ties, which are not used in the WR calculation.}

Clearly, if endpoints are censored or missing, the actual win or loss status of some pairs may not be determined properly; {thus Pocock's algorithm} may lead to a biased estimated WR. For example, in a pair with at least one missing endpoint measure, the pair is considered a tie (for the endpoint), even though the truth may be different had the data been available. Similarly, for a time-to-event endpoint subject to censoring, the win or loss depends on the common follow-up time of the paired participants; thus, we may not utilize all observed data appropriately for this pair. In particular, when one participant is censored before an event is observed for the other participant, the pair is viewed as a tie and the comparison is carried out on the subsequent endpoint. However, if there were no censoring, the tie might not have been a true tie, but an actual win or loss.  

As demonstrated by \citet{li2024elusiveness}, even under the most favorable missing data mechanism, the missing completely at random (MCAR), where missingness is independent of both observed and unobserved data \citep{rubin1976inference}, Pocock's method can yield biased estimates. 
Thus, consistent estimation of WR using Pocock's method for censored or missing data may require strong assumptions on  missing data and censoring mechanisms, arguably stronger than the standard assumptions of MCAR and non-informative, homogeneous censoring across treatment groups. 

Pocock's handling of censoring and missing data can lead to an estimator that depends on the censoring distribution or missing data mechanism, and therefore may not consistently estimate the WR. From an estimand perspective, this dependence is undesirable, since the WR should reflect treatment effects under a fixed design and study population, not the particular censoring or missingness patterns observed in a given trial. This view is consistent with the ICH E9 (R1) addendum, which emphasizes that handling missing and censored data must be an integral part of the estimand framework in clinical trials \citep{kahan2024estimands}. 

Several authors have investigated how censoring and missingness affect WR estimation and proposed potential remedies.  \citet{dong2020winimpact} showed that bias can arise when the primary endpoint is censored, particularly if treatment has long-term benefits, and highlighted censoring time and follow-up duration as key drivers of bias. Building on this observation, \citet{dong2020inverse} introduced an inverse probability of censoring weighting (IPCW) estimator that consistently estimates the WR under non-informative censoring, and later extended the method to incorporate baseline and time-varying covariates to handle covariate-dependent censoring \citep{dong2021adjusting}. \citet{mao2024defining} further noted that higher censoring rates inflate the number of ties, complicating recovery of the true WR. However, their proposed solution applies only to hierachical time-to-event endpoints. \citet{wang2025restricted} approached the problem differently, developing a joint imputation algorithm for missing longitudinal and clinical outcomes and then applying counting-based methods to estimate the WR based on a time horizon. While flexible in accommodating multiple outcome types, this approach can be computationally intensive and sensitive to misspecification of the imputation model. Except for \citet{wang2025restricted}, existing methods have focused on hierarchies composed only of survival outcomes, and have not addressed {settings} where the hierarchy also includes a non-survival endpoint, such as a cross-sectional quality-of-life (QOL) or response score. 


In this paper, we focus on a specific and practically important setting: two hierarchically ordered endpoints, where the first is a terminal time-to-event outcome subject to censoring, and the second is a non-survival endpoint that may be missing. This structure is common in trials where survival is prioritized but supplemented by patient-reported or clinical outcomes. For example, HEART-FID prioritized time to death within 12 months, followed by 6-minute walk distance at 12 months. Similarly, ISCHEMIA considered time to death within 4 years followed by the Seattle Angina Questionnaire–Angina Frequency Score (SAQ7-AF) at year 4. While our method is tailored to this structure, the framework can be extended more broadly.

We introduce a one-dimensional summary, the \textit{$\mc S$-score,} that preserves the hierarchical ordering of the two endpoints while enabling nonparametric estimation. By expressing WR as a function of the distribution of the $\mc S$-score, we obtain a plug-in estimator that accommodates both censoring and missingness without requiring strong modeling assumptions. Without the need of pairwise comparisons, our method avoids imposing strong parametric assumptions when censoring or missing data are present. Framing the problem in this way enables us to use the general theory of M-estimation, which in turn provides a principled basis for deriving influence-function–based variance estimators. Therefore, we are able to derive a close-form asymptotic variance estimator for the proposed WR estimator.

 
By avoiding issues associated with multiple pairwise comparisons or the need to specify an imputation model, 
our proposed $\mc S$-score method provides a simple nonparametric maximum likelihood estimator (NPMLE), which 
is robust to both non-informative censoring and missing at random (MAR) mechanisms. 
The $\mc S$-score method is distribution-based; it fundamentally differs from counting-based approaches such as Pocock's algorithm.  To facilitate and promote its use, we have also built a user-friendly R package \texttt{WinRS} is freely available at: \url{https://github.com/yiliu1998/WinRS}.
{Our methodological framework is developed for randomized clinical studies with two prioritized endpoints: a first-prioritized survival endpoint and a second-prioritized continuous or ordinal endpoint. It is directly applicable under non-informative censoring for the first endpoint that is independent of covariates, and under a missing-at-random mechanism for the second endpoint, where missingness may depend on observed baseline covariates. }

The remainder of this paper is organized as follows. In Section \ref{sec:method}, we present the statistical preliminaries and our proposed method. We conduct extensive simulation studies, in Section \ref{sec:simu}, to examine and showcase the finite-sample performance of our proposed estimator. In Section \ref{sec:data}, we demonstrate the utility of our framework in two case studies: the HEART-FID and ISCHEMIA clinical trials. Finally, in Section \ref{sec:concludes}, we conclude the paper with comments and remarks. {We acknowledge the use of ChatGPT-5.2 for language polishing and grammar editing, with no other uses of any large language models in the preparation of this manuscript.}

\section{Methodology}\label{sec:method}

\subsection{Notation and set-up}\label{subsec:setup}

Consider a randomized study with two treatment groups, $a$ and $b$, of sizes $n_a$ and $n_b$, respectively. We focus on a setting where we have  two hierarchical endpoints. The first endpoint, $Y_1$, is a terminal time-to-event outcome, subject to censoring, measured over an interval $[0, h]$, where $h$ denotes the finite time horizon of the study such that all participants with $Y_1$ occurring after $h$ are considered to be administratively censored. The second endpoint, $Y_2$, is measured at the end of the study (i.e., {time $h$}). 
{Note that in practice, participants usually complete follow-up in a short time window around $h$, where the value of $h$ is assumed to be the same. Therefore, measurement of $Y_2$ is treated to have occurred at time $h$.} Measures of $Y_2$ can be missing for some participants. We assume values of $Y_2$ lie within a bounded range $[0, \tau]$, for some finite $\tau > 0$, or, they can be transformed to fall within $[0, \tau]$ via an appropriate monotonic {transformation such that} both $Y_1$ and $Y_2$ are positive-valued, with larger values indicating better outcomes.

Note that $Y_2$ cannot be measured if the terminal event occurs before the follow-up visit. Therefore, we do not consider the hypothetical existence of a participant's $Y_2$ if their $Y_1$ is observed within the study period $[0, h].$ This is akin to competing risk scenarios \citep{gooley1999estimation, fine1999proportional, putter2007tutorial} or multi-state models \citep{hougaard1999multi}, where $Y_2$ may not be well-defined once the terminal event $Y_1$ has occurred as the occurrence of $Y_1$ precludes the possibility to measure, observe, or experience $Y_2$. 

\subsection{The win ratio estimand and conventional estimation approach}\label{subsec:wr-estimand} 
For time-to-event endpoint $Y_1$ that is observable in study period $[0, h]$, there is a corresponding event time $T_1$, where $T_1\in(0,\infty)$ such that
$$
Y_1=\min(T_1,h) + I(T_1>h), 
$$
where $h$ denotes the study's time horizon and $I(\cdot)$ is the indicator function. This formulation yields a coarsened version of $T_1$, representing the practical scenario in which the event may have not been occurred by the end of the study. For convenience,  we record $Y_1 = h + 1$ if the terminal event occurs after time horizon $h$. $Y_1$ can be right-censored, a type of missing data, if censoring occurs before $h$.  

The estimand of interest is the WR, defined based on two hierarchical endpoints with pre-specified time horizon or follow-up time $h < \infty$. This WR is well-defined to make the follow-up time explicit and corresponds to the ``restricted time WR'' estimand considered in \citet{wang2025restricted}. Let $Y_{1z}$ and $Y_{2z}$ denote the  first and second hierarchical endpoints, respectively, for group $z \in \{a, b\}$. For a random pair of participants, one from each group,  if $Y_{1a} > Y_{1b}$, then group $a$ wins and  conversely, if $Y_{1a} < Y_{1b}$, group $a$ loses, based on the first endpoint. When they are tied with $Y_{1a} = Y_{1b}$, the second endpoint is used for comparison: if $Y_{2a} > Y_{2b}$, group $a$ wins, and if $Y_{2a} < Y_{2b}$, group $a$ loses. A tie occurs only when they are tied on both endpoints, i.e., ($Y_{1a} = Y_{1b}$ and $Y_{2a} = Y_{2b}$). The WR of interest is defined as the ratio of the probability of wins by group $a$ to the probability of loses by group $a$, given by
\begin{align}\label{eq:WR-estimand}
\WR &= \frac{P(Y_{1a} > Y_{1b}) + P(Y_{1a} = Y_{1b}, Y_{2a} > Y_{2b})}{P(Y_{1a} < Y_{1b}) + P(Y_{1a} = Y_{1b}, Y_{2a} < Y_{2b})}.
\end{align}
This estimand provides a practical, meaningful, and interpretable treatment effect at time horizon $h$ based on the two hierarchical endpoints. While this esimand depends on the time horizon, but it does not depend on the censoring distribution of $Y_1$ nor missing data mechanism for $Y_2$.




A consistent estimator of this WR estimand for data without censoring nor missingness can be obtained via Pocock's pairwise comparison algorithm \citep{pocock2012win}. We briefly review it as follows. Each participant in one of the two treatment groups is paired with each of all participants in the other group, which results in $n_a\times n_b$ possible different pairs. {The first endpoint on the hierarchy is used for comparison first between  each pair: the participant in group a is declared a ``winner'' if this participant fares better relative to the other participant in group $b$; the pair is then classified as a ``win'' for group $a$}. If the pair is tied on the first endpoint, the algorithm then proceeds to compare the pair on the next endpoint in the hierarchy, and so forth, until all pairs are declared to be winners, losers or ties for group $a$. The WR is then estimated by
\begin{align}\label{eq:WR-pocock}
    \WRhat=\frac{\text{Total number of wins in group }a}{\text{Total number of loses in group }a}. 
\end{align}
The variance of Pocock's estimator is usually based on U-statistics framework developed by \citet{bebu2016large}, allowing for the construction of Wald-type confidence intervals. Alternatively, other robust variance estimation approaches are also available, such as those proposed by \citet{matsouaka2022robust}.

In the following section, we propose a simple approach for estimating the WR under our setting of two hierarchical endpoints. 

\subsection{Estimation method in absence of censoring or missing data}\label{subsec:propose-full}
Suppose we do not have censoring nor missing data.  Consider the following score $\mc S$, a new random variable that  integrates  information from the two endpoints:
\begin{align}\label{eq:Sscore}
    \mc S=Y_1 + I(Y_1 > h)Y_2.
\end{align}
We have the following equivalence as shown in Appendix \ref{subapp:proof-equi}. 
\begin{align}\label{thm:equi}
\{\mc S_a > \mc S_b\} = \{Y_{1a} > Y_{1b}\} \cup \{Y_{2a} > Y_{2b}, Y_{1a} = Y_{1b}\} \quad \text{with probability } 1.
\end{align}

This equivalence relies on the fact that, in our setting, the ability to observe $Y_2$ at time $h$ implies that $Y_1 > h$, as $Y_1$ is time to a terminal event (see Section \ref{subsec:setup}). Consequently, $\{Y_{2a} > Y_{2b}, Y_{1a} = Y_{1b} \leq h\}$ cannot occur and has zero probability.

\begin{remark}\label{rmk:cont-time}
   For any fixed time point $k \leq h$, the probabilities $P(Y_{2a} > Y_{2b}, Y_{1a} = Y_{1b} = k)$ and $P(Y_{2a} < Y_{2b}, Y_{1a} = Y_{1b} = k)$ are both zero if the event time $T_1$ is continuously distributed over $[0, h]$, since $\{Y_{1a} = Y_{1b} = k\}$ is a zero-probability set. As a result, the event $\{Y_{2a} > Y_{2b}, Y_{1a} = Y_{1b} \leq h\}$ also has probability zero---regardless of whether $Y_1$ is a terminal event. This implies that, under the continuous time setting, the condition that $Y_1$ is terminal can be relaxed to imply Equation \eqref{thm:equi}, thereby broadening the applicability of our method.  
\end{remark}

With the equivalence in \eqref{thm:equi}, we have $P(\mc S_a>\mc S_b)=P(Y_{1a}>Y_{1b}) + P(Y_{2a}>Y_{2b}, Y_{1a}=Y_{1b})$. This helps to re-write the WR estimand defined in \eqref{eq:WR-estimand} to be 
\begin{align}\label{eq:eqWR}
    \WR=\frac{P(\mc S_a>\mc S_b)}{P(\mc S_b>\mc S_a)}=\frac{\displaystyle\int\bar F_a(s)d F_b(s)}{\displaystyle\int\bar F_b(s)d F_a(s)},
\end{align}
where $F_z$ and $\bar F_z$ are, respectively, the cumulative distribution function (CDF) and survival function of the $\mc S$-score under group $z$, for $z\in\{a,b\}$. Therefore, estimating the WR can be based on the estimated distributions of the $\mc S$-scores for the two treatment groups. We refer to our proposed method as the ``$\mc S$-score method'' or ``$\mc S$-score estimator'' in the remainder of the paper. 


Based on Equation \eqref{thm:equi}, the condition $\mc S_a > \mc S_b$ serves as the criterion for a ``win'' of group $a$. Thus, wining or losing based on hierarchical endpoints $Y_1$ and $Y_2$ is simplified to winning of losing based on one endpoint of $\mc S$. The support of $\mc S$ is given by $(0, h+\tau+1]$.  Note that the cdf and survival function of $\mc S$ have a jump in interval $[h, h+1]$ since $Y_1$ has a jump from $h$ to $h+1$ by definition. Thus, we define that, for all $s\in[h, h+1)$, $F_z(s)=F_z(h)$ for $z\in\{a,b\}$. A simple NPMLE for $\WR$ based on full data $\mc O^{\text{full}}= (Z, Y_1, Y_2)$ with $Z$ as the treatment group is
\begin{align}\label{eq:WR-npmle}
    \WRhat(\mc O^{\text{full}}) = \frac{\widehat P(\mc S_a > \mc S_b)}{\widehat P(\mc S_b > \mc S_a)}=\frac{\displaystyle\int_0^{h+\tau+1} \{1-\widehat F_a(s)\} d\widehat F_b(s)}{\displaystyle\int_0^{h+\tau+1} \{1-\widehat F_b(s)\} d\widehat F_a(s)}, 
\end{align}
where $\widehat F_a$ and $\widehat F_b$ are the Kaplan-Meier estimates for the CDF 
\citep{kaplan1958nonparametric} by treating $\mc S$ as a time-to-event variable where all participants have an event at their observed value of $\mc S$. Specifically,  
\begin{align}
    \widehat F_z(s) = 1 - \prod_{s_i^{(z)} \leq s} \left(1 - \frac{d_i^{(z)}}{n_i^{(z)}}\right),
\end{align}
where $s_i^{(z)}$ denotes the distinct observed value of ${\mc S}$ that are treated as ``event time'' from data indexed by $i$, $d_i^{(z)}$ represents the number of events occurring at the time $s_i^{(z)}$, and $n_i^{(z)}$ is the number of individuals at risk just prior to time $s_i^{(z)}$ in group $z$, for $z\in\{a,b\}$. 

The proposed estimator \eqref{eq:WR-npmle} 
differs from Pocock's estimator \eqref{eq:WR-pocock} in the sense that it is distribution-based rather than a counting-based approach. This allows the use of M-estimation theory to obtain influence-function-based variance estimator. 

Thus far, our proposed estimator is for fully observed data $\mc O^{\text{full}}$. In practice, however, a more realistic scenario involves censoring in $Y_1$ before time horizon $h$ and missing data in $Y_2$ at the follow-up time. To address this, Pocock's method deals with censoring by using common follow-up time of the paired participants for pairwise comparison in $Y_1$ and considers the paired participants to be ties in $Y_2$ if one of them has missing data. {This approach can yield biased estimates because the win/lose/tie classification based on the common follow-up time ($< h$) in $Y_1$ due to censoring may differ from the classification that would have been obtained if both participants had been observed up to time $h$. In addition, two participants who appear tied in $Y_2$ under missing data may not be tied if both $Y_2$ values were fully observed.} 
Pocock's method may implicitly rely on strong assumptions about the censoring and missingness mechanisms that may be stronger than the standard assumptions of non-informative censoring for $Y_1$ and MCAR for $Y_2$ \citep{li2024elusiveness, mao2024defining}. While the independent censoring assumption for $Y_1$ is widely accepted in survival analysis \citep{cox1972regression, kaplan1958nonparametric}, especially in randomized studies, the MCAR assumption for $Y_2$ is often unrealistic.  As noted by \citet{mao2024defining}, Pocock's method is sensitive to the censoring rate of the survival endpoint. 

In the following section, we extend the proposed $\mc S$-score method to estimate WR for our setting with censoring data in $Y_1$ and missing data in $Y_2$.

\subsection{Estimation method when there is censoring or missing data}\label{subsec:propose-obs}

Suppose that $Y_1$ is right-censored before $h$ with censoring random variable $C_1$ and let 
$$
\widetilde{Y}_1=\min(Y_1, C_1, h) + I(Y_1 > h, C_1\geq h) \text{ with } \Delta_1=I(Y_1\leq C_1, Y_1\leq h) + I(Y_1 > h, C_1\geq h).
$$ 
The observed value of $Y_1$ can be represented by $(\widetilde{Y}_1, \Delta_1),$ with $\Delta_1 = 0$ indicating right-censoring within the interval $[0, h]$. Note that $Y_1 = h + 1$ represents administrative censoring due to time horizon of $h$ and is not considered as censoring here. For missing data in $Y_2$, let $R_2$  be the non-missingness indicator, i.e., $R_2=1$ when $Y_2$ is observed and $R_2=0$ otherwise. Note that $Y_2$ is naturally missing if the terminal event occurred before $h$. 

We denote the coarsening-level version of the full data as $\mc O=(Z, \widetilde Y_1, \Delta_1, \widetilde Y_2, R_2)$, where $\widetilde Y_2=Y_2R_2$ and we make the following two assumptions. 

\begin{assumption}[Non-informative censoring]\label{asp:censor}
The censoring is non-informative within each treatment group in the sense of $C_1\bigCI(Y_1, Y_2)\mid Z=z$,
where $\bigCI$ denotes ``independence.'' 
\end{assumption}

\begin{assumption}[Missing at random (MAR)]\label{asp:MAR} For the non-survival endpoint $Y_2$, missing data are missing at random after time horizon $h$, 
i.e., $Y_2\bigCI R_2\mid Y_1>h, Z=z$. 
\end{assumption}
The independence between censoring $C_1$ and event times $Y_1$ is commonly assumed in the survival analysis literature \citep{kaplan1958nonparametric, cox1972regression} to ensure the identifiability of the event's survival function based on the nonparametric Kaplan-Meier estimator. To be more general, in Assumption \ref{asp:censor}, we allow the censoring mechanism to differ under each treatment group. The reason for including $C_1 \bigCI Y_2 \mid Z=z$ in Assumption \ref{asp:censor} is to ensure that samples with $Y_1>h$ remain representative for estimating the distribution of the $\mc S$-score beyond $h$. For instance, when $C_1<h$ but $Y_1>h$, $Y_2$ could exist but remain unobserved. If $Y_2$ were dependent on $C_1$, this would introduce bias when estimating the CDF of the $\mc S$-score for $s \geq h$. 



We also formulate the MAR assumption (Assumption \ref{asp:MAR}) such that any missing data are random conditional on not having experienced the terminal event 
(condition $Y_1>h$) within the treatment group membership (condition $Z=z$). The condition $Y_1>h$ is when $Y_2$ can be measured at the follow-up time.

Assumption \ref{asp:MAR} inherently aligns with the concept of ``monotone missing data'' in the study design, which refers to a missing data pattern where the occurrence of missing data for a variable implies that all subsequent variables (in a predefined order) are also missing \citep{tsiatis2007semiparametric}. In our context, when $Y_1$ is censored (also a type of missingness) before $h$, $Y_2$ is automatically missing (under a mechanism that is independent of the censoring of $Y_1$ by Assumption \ref{asp:censor}). Therefore, we discuss missing data patterns of $Y_2$ only for participants with $Y_1 > h$. 

\begin{figure}[H]
    \centering
    \includegraphics[width=\linewidth]{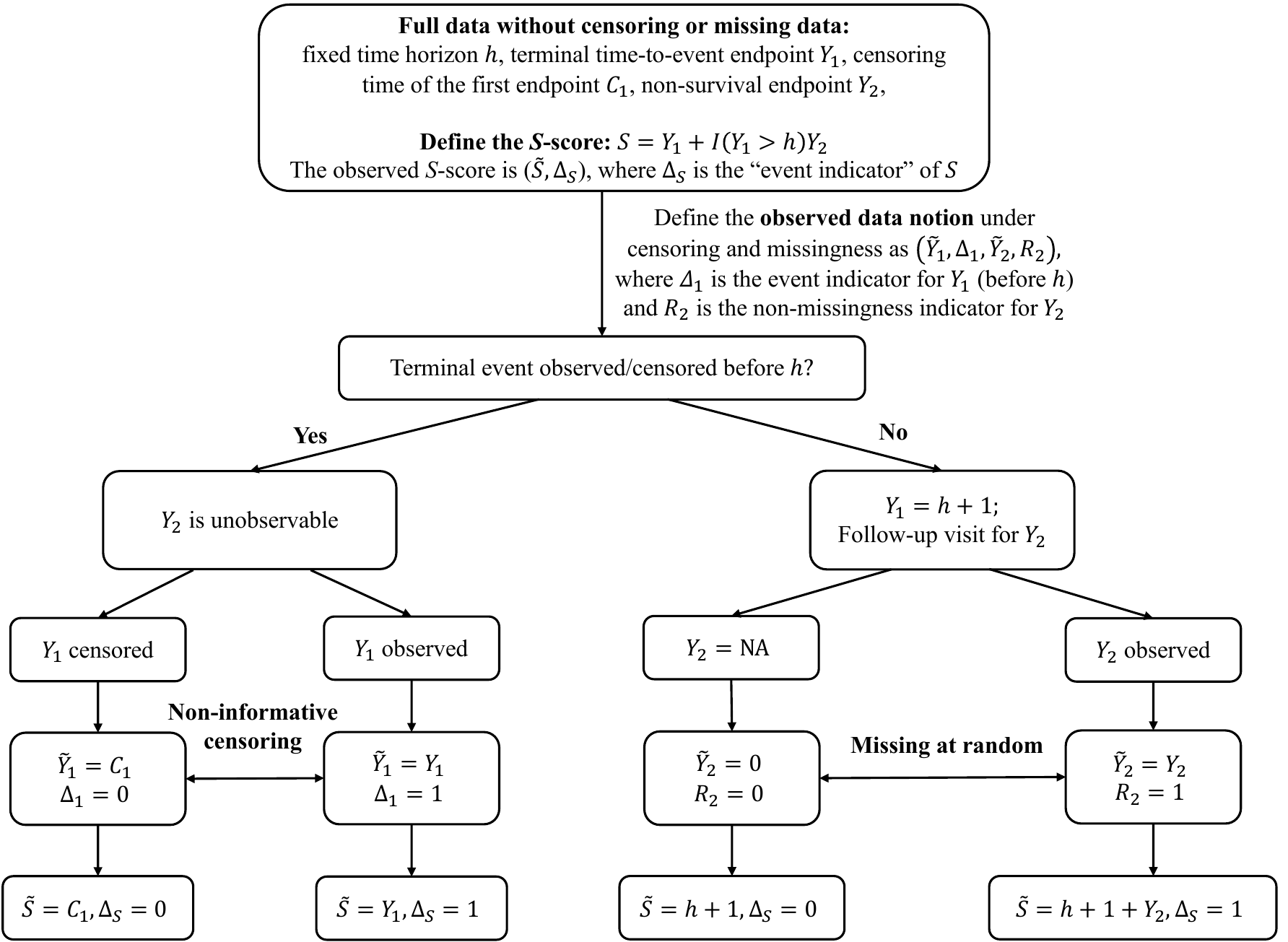}
    \caption{{An illustration flow of the proposed $\mc S$-score methodology.}}
    \label{fig:flow}
\end{figure}

Based on the observed data, we extend our $\mc S$-score defined in \eqref{eq:Sscore} from the full data $\mc O^{\text{full}}$ to $\mc O$. We introduce the following coarsened $\mc S$ as a ``right-censored'' survival endpoint. Denote $\widetilde{\mc S}=\widetilde Y_1 + I(\widetilde Y_1>h)\widetilde Y_2$ and the ``event'' indicator by $\Delta_{\mc S}=I(\Delta_1=1, \widetilde Y_1\leq h) + I(R_2=1, \widetilde Y_1=h+1)$. Notice that with probability 1, $\{\Delta_1=1, \widetilde Y_1\leq h\}$ and $\{R_2=1, \widetilde Y_1=h+1\}$ are disjoint events, and thus $\Delta_{\mc S}\in\{0,1\}$. Therefore, our observed data can be re-written as $\mc O^{\mc S}=(Z, \widetilde{\mc S},\Delta_{\mc S})$. {We summarize the transformation from the two endpoints to the $\mc S$-score in Figure \ref{fig:flow}.}

Note that $\widetilde{\mc S}$ is the observed version of ${\mc S}$ with ``censoring'' of $\mc S$ based on: (i) the right-censoring status of $Y_1$, and (ii) the missingness of $Y_2$. Under Assumptions \ref{asp:censor} and Assumption \ref{asp:MAR} , the censoring is non-informative. Thus, the Kaplan-Meier estimators based on $\mc O^{\mc S}$ for each group are unbiased for the survival functions of $\mc S$ for each group. Thus, we have the following simple NPMLE for WR for data with censoring and missingness:
\begin{align}\label{eq:WR-npmle-obs}
    \WRhat(\mc O^{\mc S}) = \frac{\widehat P(\mc S_a > \mc S_b)}{\widehat P(\mc S_b > \mc S_a)}=\frac{\displaystyle\int_0^{h+\tau+1} \{1-\widehat F_a(s)\} d\widehat F_b(s)}{\displaystyle\int_0^{h+\tau+1} \{1-\widehat F_b(s)\} d\widehat F_a(s)}, 
\end{align}
where $\widehat F_a$ and $\widehat F_b$ are the Kaplan-Meier estimates for the CDF 
of $\mc S$ based on $\mc O^{\mc S}$ for each group.  
This estimator in Equation \eqref{eq:WR-npmle-obs} is shown to be consistent for $\WR$ in Appendix \ref{subapp:proof-consistency}.

\begin{remark}\label{rmk:Cs}
    We have made an implicit assumption of non-informative censoring for the $\mc S$-score by Assumptions \ref{asp:censor} and \ref{asp:MAR} as well as the definitions of $\widetilde{\mc S}$ and $\Delta_{\mc S}$.  Specifically, we can define a ``censoring'' variable $C_{\mc S}$ for the $\mc S$-score as follows:
    \begin{align*}
      C_{\mc S} = \begin{cases}
      C_1, & \quad \text{if }C_1<h, \\
      h+0.5,  & \quad \text{if }C_1\geq h\text{ and }R_2=0, \\
      h + \tau + 1 + 0.5, &  \quad \text{if }C_1\geq h\text{ and }R_2=1. 
    \end{cases}
    \end{align*}
Under Assumptions \ref{asp:censor} and \ref{asp:MAR}, $\mc S\bigCI C_{\mc S}\mid Z=z$. 
Thus, the Kaplan-Meier estimator based on variables $\widetilde{\mc S} = \min(\mc S, C_{\mc S})$ with $\Delta_{\mc S} = I(\widetilde{\mc S}<C_{\mc S})$ can be used in \eqref{eq:WR-npmle-obs}.
\end{remark}


\subsection{Variance estimation}\label{subsec:varest}

To be able to carry out the inference for the proposed NPMLE \eqref{eq:WR-npmle-obs}, we need an appropriate variance estimator for the estimator \eqref{eq:WR-npmle}. A commonly-used approach is the nonparametric bootstrap \citep{efron1994introduction}, known for its effectiveness and accuracy. Given that the proposed estimator \eqref{eq:WR-npmle} is regular and asymptotically linear (RAL) and smooth with respect to the $\mc S$-score with probability 1 except at certain jump points in the survival functions, the bootstrap method provides accurate variance estimation, as justified by \cite{shao2012jackknife}. The confidence interval (CI) via bootstrap can be constructed in multiple ways. For example, one can use the estimated variance from multiple bootstrap samples to obtain a Wald-type CI. Alternatively, the CI can be constructed using the $\alpha/2$ and $(1-\alpha/2)$ quantiles of the bootstrap distribution as bounds. 

However, a notable drawback of the bootstrap is its computational burden, which increases as the sample size increases. Therefore, it is a common practice to leverage asymptotic theory to derive a closed-form analytical variance estimator alongside the point estimator, providing there is an alternative for a quick estimation for the uncertainty. In this paper, we leverage the theory of influence function (IF) \citep{van2000asymptotic, stefanski2002calculus} to propose an asymptotic variance estimator for \eqref{eq:WR-npmle}. The IF, denoted by $\varphi(\mc O^{\mc S};F_a,F_b)$, depends on the observed data and the CDFs of $\mc S$-score for both treatment groups. The formula of the IF of \eqref{eq:WR-npmle} is derived in Appendix \ref{subapp:varest}. Moreover, the  influence function $\varphi(\mc O^{\mc S};F_a,F_b)$ is such that
\begin{align*}
    \WRhat = \WRhat(\mc O^{\mc S}) = \WR+\frac1{\sqrt{n}}\sum_{i=1}^n\varphi(\mc O^{\mc S}_i;  F_a, F_b) + o_p(n^{-1/2}). 
\end{align*} This allows us to use the standard central limit theorem to approximate the asymptotic variance as
\begin{align*}
    \widehat{\mc V}_{\text{IF}} = \widehat{\mc V}_{\text{IF}}(\WRhat) = \frac1n\sum_{i=1}^n \varphi(\mc O^{\mc S}_i; \widehat F_a, \widehat F_b)^2. 
\end{align*}
In other words, we can show the following consistency and asymptotic normality (see Appendix \ref{subapp:varest}), 
\begin{align*}
    \sqrt{n/\widehat{\mc V}_{\text{IF}}}(\WRhat-\WR)\xrightarrow{d}\mc N(0,1). 
\end{align*}

\subsection{Extension to incorporate baseline covariates}\label{subsec:covar}

In this section, we extend  Assumption \ref{asp:MAR} to allow the missing data mechanism to depend on the baseline covariates, in addition to the group indicator. Using  a vector of baseline covariates $\mb X$, we make the following assumption:
\begin{assumption}[Missing at random (MAR) with covariates]\label{asp:MAR-X}
    $R_2\bigCI Y_2\mid Y_1 > h, Z=z, \mb X$. 
\end{assumption}

Assumption \ref{asp:MAR-X} generalizes Assumption \ref{asp:MAR}, as it allows individual-specific missingness patterns. Under Assumption \ref{asp:MAR-X}, covariate adjustment is required for consistent estimation. By contrast, if Assumption \ref{asp:MAR} holds, consistent estimation can be achieved without covariate adjustment. However, modeling missing data with covariates may help gaining efficiency \citep{liu2025coadvise, gao2024does}. 

Under Assumptions \ref{asp:censor} and \ref{asp:MAR-X}, the WR can be estimated by solving a set of unbiased estimating equations, by using the M-estimation theory \citep{van2000asymptotic}. Specifically, we construct a stacked set of estimating equations incorporating the Kaplan–Meier estimator for the CDF of $Y_1$, the treatment-specific regression model parameters for the missing data mechanism of $Y_2$, and the covariate-adjusted CDF estimator for $Y_2$. Estimation of both the missingness mechanism and the CDF of $Y_2$ is restricted to the subset of data with $Y_1 > h$. 
Full technical details on the estimating equations, the resulting covariate-adjusted point estimator, and its asymptotic properties are provided in Appendix \ref{subapp:coad}. 

While we refer to the equation \eqref{eq:WR-npmle-obs} as the NPMLE, the covariate-adjusted version is more accurately described as a semiparametric estimator, as it incorporates a parametric model for the missingness mechanism, parameterized by a vector $\bd\beta_z$ (of the same dimension as $\mb X$) for $z = a,b$, essentially a model with main effects of covariates, treatment indicator, and interaction effects of baseline covariates and treatment indicator. Analogous to the variance estimator of the NPMLE in Section \ref{subsec:varest}, we derive an IF-based variance estimator for the covariate-adjusted semiparametric estimator (see Appendix \ref{subapp:coad}). This covariate-adjusted estimator, along with its IF-based variance estimator, is also implemented in our R package \texttt{WinRS}. 


\section{Simulation Study}\label{sec:simu}

We conducted simulation studies to compare our proposed method with the methods by \cite{pocock2012win} and  \cite{wang2025restricted}{, as well as the IPCW method (\cite{dong2020inverse}; only for the survival endpoint),} which we simply labeled as ``Pocock'', ``Wang'' {and ``IPCW''}, respectively. For the data-generating process described in Section~\ref{subsec:DGP} below, we simulated two sets of Monte Carlo data. First, we generated full data $\mc O^{\text{full}, z} = (Z=z, Y_1, Y_2)$ for $z \in \{a, b\}$ under a super-population of size $n_{\text{super}} = 10^7$ per group. From the combined pool $\mc O^{\text{full}, a} \cup \mc O^{\text{full}, b}$, we randomly sampled $10^7$ independent pairs, each consisting of one participant from each group. The WR was then computed based on pairwise comparisons across these pairs by counting the number of wins for groups $a$ and $b$. Given the large sample size, number of comparisons, and complete data (i.e., with no censoring or missing data), the uncertainty in estimating the true WR is negligible. This large sample allows us to calculate the ``true'' WR, which we will use to evaluate performance measures needed to compare different estimation methods using the second Monte Carlo data below.

Second, we simulated $2,000$ independent observed datasets $\mc O^z = (Z=z, \widetilde Y_1, \Delta_1, \widetilde Y_2, R_2)$ for $z \in \{a, b\}$, under randomized allocation with a 1:1 ratio and varying sample sizes $(n_a, n_b) \in \{(100, 100), (1000, 1000)\}$ to represent small and large trials. For each replicate, we estimated the WR using {the four methods aforementioned (Pocock, Wang, IPCW and our $\mc S$-score).} 

Pocock's method is implemented via the \texttt{WinRatio} R package \citep{duarte2022winratio}, using the default unmatched-pair algorithm and its built-in CI estimator.

We followed the instructions in Appendix A.1 of \cite{wang2025restricted} for multiple imputations on right-censored time-to-event and continuous endpoints 
to implement their method (imputation-then-counting). Following their simulation study framework, we chose to perform imputations over 50 imputed datasets and apply Pocock's algorithm to each imputed dataset. 

Let $\widehat\theta_i^{\text{WR}}$ and $\widehat{\mc V}_i$ denote the point and variance estimates from the $i$th imputed dataset, $i=1,\dots,50$. The coresponding point estimate and variance can be estimated by 
$$
\bar\theta^{\text{WR}} = \frac{1}{50}\sum_{i=1}^{50}\widehat\theta_i^{\text{WR}} ~~~\text{ and } ~~~\widehat{\mc V}_{\text{pooled}} = \widehat{\mc V}_{\text{within}} + \left(1+\frac{1}{50}\right)\widehat{\mc V}_{\text{between}},
$$  
where  
$
 \widehat{\mc V}_{\text{within}} = \displaystyle\frac{1}{50}\displaystyle\sum_{i=1}^{50}\widehat{\mc V}_i,  
\quad
\widehat{\mc V}_{\text{between}} = \displaystyle\frac{1}{49}\displaystyle\sum_{i=1}^{50} \left(\widehat\theta_i^{\text{WR}} - \bar\theta^{\text{WR}}\right)^2.
$\\
We then construct Wald-type confidence intervals based on $\widehat{\mc V}_{\text{pooled}}$.

{For the IPCW method of \cite{dong2020inverse}, we note that it was originally developed for settings with one or multiple censored survival endpoints. Although this setting differs from ours, the method is implemented in the \texttt{WINS} R package and can be applied in our simulations. Specifically, we implemented this approach using the \texttt{win.ratio()} function, specifying the first endpoint type as \texttt{"tte"} (time-to-event) and the second as \texttt{"continuous"}, with the estimation method set to \texttt{"ipcw"}. For the continuous endpoint, the package adopts a complete-case analysis in the presence of missing data. Since both Wang's and IPCW methods are computationally intensive, we only applied these two approaches in the scenarios presented in the main paper, but not in the scenarios presented in Online Supplemental Material (Web Appendix S.1). }

For our $\mc S$-score method, we constructed CIs for $\WR$ using two different Wald-type CIs based on our closed-form variance estimator derived from the influence function (IF-Wald, Section~\ref{subsec:varest}) and using the standard error estimated from 1,000 nonparametric bootstrap replicates (BT-Wald). We also consider a  percentile-based CI using the 2.5\% and 97.5\% quantiles of the bootstrap estimates (BT-QT).

We assessed the performance of the WR estimation methods over the 2,000 simulated datasets using the following four metrics:
\begin{itemize}
    \item Absolute relative percent bias (ARB\%): The absolute value of the mean of $(\WRhat - \WR)\% / \WR$.
    \item Root mean square error (RMSE): The square root of the mean of $(\WRhat - \WR)^2$.
    \item Coverage probability (CP\%): The percentage of the number of times the true $\WR$ falls within the constructed 95\% CIs. Due to the Monte Carlo error, a CP\% within $95 \pm 1.96 \times \sqrt{95 \times (100-95) / 2,000} = [94, 96]$ is considered not significantly different from the nominal 95\% level.
    \item Width of CI: The average width of the 95\% CIs quantifying differences among the variance estimators.
\end{itemize}

\subsection{Data generating process}\label{subsec:DGP}

We generated data under different scenarios of sample size, right censoring pattern of $Y_1$, missing data pattern of $Y_2$ as well as the treatment effect of $\WR=1$ or $2$ to comprehensively assess the proposed method. 

For each $z \in \{a, b\}$, we generated both the event times $T_{1z} \sim \Gamma(\alpha_{Tz}, \lambda_{Tz})$  and censoring times $C_{1z} \sim \Gamma(\alpha_{Cz}, \lambda_{Cz})$ from the Gamma distributions $\Gamma(\alpha, \lambda)$, with $\alpha$ and $\lambda$ the shape and scale parameters, respectively. 
With $h = 90$, we then generated the observed time as $Y_{1z} = \min(T_{1z}, h) + I(T_{1z} > h)$ and the event indicator as $\Delta_{1z} = I(Y_{1z} \leq C_{1z}, Y_{1z} \leq h) + I(Y_{1z} > h,C_{1z} > h)$. 

With $\tau = \max(Y_2)=50$, we also generated $Y_{2z}\sim I(Y_{1z}>h)\cdot\min\{\max(\mc N(\mu_z,\sigma_z^2), 0), \tau\}$
and the missingness indicator as
$R_{2z}^* \sim I(T_{1z} \leq h) + I(T_{1z} > h) \cdot \text{Bernoulli}(\beta_z)$ (so the non-missingness indicator is $R_{2z}=1-R_{2z}^*)$. That is, we only generated $Y_2$ for those with $Y_1>h$ since $Y_1$ is a terminal event and we can only measure $Y_2$ after $h$, and $Y_2$ may be missing depending on the value of $\beta_z$. 

The following parameter values were used for different simulation scenarios:
\begin{itemize}
    \item Treatment effect: $\WR=1$ for $\alpha_{Ta}=\alpha_{Tb}=2.5$, $\lambda_{Ta}=\lambda_{Tb}=0.04$, $\mu_a=\mu_b=10$, and $\sigma_a=\sigma_b=10$; as well as $\WR=2$ when $\alpha_{Ta}=2.5, \alpha_{Tb}=4$, $\lambda_{Ta}=0.04, \lambda_{Tb}=0.10$, $\mu_a=10, \mu_b=20$, $\sigma_a=10$, and $\sigma_b=20$. 
    \item Right censoring pattern for $Y_1$:
    \begin{itemize}
        \item No censored data:  $C_1\equiv h+1$ (that is always greater than $Y_1$) in the code;
        \item Homogeneous censoring by treatment group:  $\alpha_{Ca}=\alpha_{Cb}=1.8$, $\lambda_{Ca}=\lambda_{Cb}=0.01$ (resp. 0.02) for low (resp. moderate) censoring, yielding about 20\% (resp. 40\%) censored observations in both groups overall; 
        \item Heterogeneous censoring by treatment group:  $(\alpha_{Ca}, \alpha_{Cb})=(3.2,1.5)$, $(\lambda_{Ca},\lambda_{Cb})=(0.04, 0.08)$ (resp. $(0.02, 0.05)$ for low (resp. moderate) censoring, leading to overall 20\% (resp. 40\%) censored observations in both groups overall, but each group has different censoring rates. 
    \end{itemize}
    \item Missingness pattern for $Y_2$: \begin{itemize}
        \item No missing data:  $\beta_a=\beta_b=0$; 
        \item MCAR: we consider two cases $(\beta_a,\beta_b)=(0.2, 0.2)$ and $(0.4,0.4)$ which results in about 20\% and 40\% missing data in $Y_2$, respectively, and the missing data mechanism is the same for both groups; 
        \item MAR: we consider $(\beta_a,\beta_b)=(0.15, 0.25)$ and $(0.3,0.5)$, yielding overall about 20\% and 40\% missing data in both groups, respectively. However, in these cases, treated and control groups have different proportions of missing data. That is, the missing data mechanism is dependent to the treatment assignment, which is not completely at random. 
    \end{itemize}
\end{itemize}
Overall, our experiment considered $2\times 5\times 5\times 2=100$ scenarios to assess the performance of three competing methods for point estimation ($\mc S$-score, Pocock, Wang {and IPCW}) and three CI estimators (IF-Wald, BT-Wald and BT-QT) for the $\mc S$-score method. 

\subsection{Results}\label{subsec:results}

\begin{table}
\begin{threeparttable}
    \centering
    \scriptsize
    \begin{tabular}{cclrrrrrrrrrrrrrrrrr}
    \toprule
    & & & \multicolumn{4}{c}{$\WR=1$} & \multicolumn{4}{c}{$\WR=2$}\\
    \makecell[c]{Censoring \\ of $Y_1$} & \makecell[c]{Missingness \\ of $Y_2$} & Method  & ARB\% & RMSE & CP\% & Width & ARB\% & RMSE & CP\% & Width \\
    \cmidrule(lr){1-3}\cmidrule(lr){4-7}\cmidrule(lr){8-11}
   &  & $\mc S$-score (IF-Wald) & 0.37 & 0.052 & 95.00 & 0.20 & 0.30 & 0.114 & 95.00 & 0.45 \\ 
   & No & $\mc S$-score (BT-Wald) & 0.37 & 0.052 & 94.50 & 0.20 & 0.30 & 0.114 & 95.05 & 0.45 \\ 
   &  & $\mc S$-score (BT-QT) & 0.37 & 0.052 & 94.00 & 0.20 & 0.30 & 0.114 & 94.35 & 0.45 \\ 
   &  & Pocock & 0.37 & 0.052 & 94.65 & 0.20 & 0.30 & 0.114 & 94.90 & 0.45 \\ 
  &  &  IPCW$^*$ &  0.37 &   0.052 &  \textbf{96.20} &   0.20 &   0.30 &  0.114 &  \textbf{96.60} &  0.47 \\ 
   \addlinespace
   &  & $\mc S$-score (IF-Wald) & 0.10 & 0.052 & 95.05 & 0.20 & 0.01 & 0.113 & 95.45 & 0.45 \\ 
 No censored $Y_1$ & MCAR & $\mc S$-score (BT-Wald) & 0.10 & 0.052 & 94.80 & 0.20 & 0.01 & 0.113 & 95.50 & 0.45 \\ 
   & (40\%) & $\mc S$-score (BT-QT) & 0.10 & 0.052 & 94.50 & 0.20 & 0.01 & 0.113 & 95.00 & 0.45 \\ 
   &  & Pocock & 0.11 & 0.054 & 95.45 & 0.21 & 9.67 & 0.226 & \textbf{53.30} & 0.41 \\ 
   & & Wang & 0.11 & 0.053 & 94.55 & 0.20 & 0.08 & 0.113 & 95.30 & 0.45 \\
   &  &  IPCW$^*$ &  0.28 &   0.052 &  96.00 &   0.21 &   9.66 &  0.227 &  \textbf{57.60} &  0.43 \\ 
   \addlinespace
   &  & $\mc S$-score (IF-Wald) & 0.48 & 0.051 & \textbf{96.40} & 0.20 & 0.08 & 0.110 & 95.90 & 0.45 \\ 
   & MAR & $\mc S$-score (BT-Wald) & 0.48 & 0.051 & \textbf{96.40} & 0.20 & 0.08 & 0.110 & 95.85 & 0.45 \\ 
   & (40\%) & $\mc S$-score (BT-QT) & 0.48 & 0.051 & 95.85 & 0.20 & 0.08 & 0.110 & 95.70 & 0.45 \\ 
   &  & Pocock & 14.57 & 0.152 & \textbf{15.95} & 0.18 & 16.80 & 0.362 & \textbf{8.70} & 0.38 \\ 
   & & Wang & 0.89 & 0.052 & \textbf{96.30} & 0.20 & 0.11 & 0.115 & 95.90 & 0.45 \\
   &  &  IPCW$^*$ &  14.11 &   0.148 &  \textbf{19.00} &  0.18 &  16.99 &  0.367 &  \textbf{10.80} &  0.39\\ 
   \midrule
   &  & $\mc S$-score (IF-Wald) & 0.27 & 0.061 & 94.15 & 0.24 & 0.14 & 0.131 & 94.15 & 0.50 \\ 
   & No & $\mc S$-score (BT-Wald) & 0.27 & 0.061 & 94.30 & 0.24 & 0.14 & 0.131 & 94.05 & 0.50 \\ 
   &  & $\mc S$-score (BT-QT) & 0.27 & 0.061 & 94.20 & 0.23 & 0.14 & 0.131 & \textbf{93.85} & 0.49 \\ 
   &  & Pocock & 0.33 & 0.068 & 94.55 & 0.26 & 2.70 & 0.152 & \textbf{92.05} & 0.53 \\ 
   & & Wang & 0.37 & 0.063 & \textbf{93.95} & 0.24 & 6.39 & 0.183 & \textbf{76.40} & 0.47 \\
   &  &  IPCW$^*$ &  0.32 &   0.071 &  \textbf{93.60} &   0.26 &  2.77 &  0.143 &  94.60 &  0.54 \\ 
    \addlinespace 
   &  & $\mc S$-score (IF-Wald) & 0.03 & 0.059 & 94.70 & 0.24 & 0.16 & 0.126 & 94.80 & 0.50 \\ 
  40\% censored $Y_1$, & MCAR & $\mc S$-score (BT-Wald) & 0.03 & 0.059 & 94.70 & 0.24 & 0.16 & 0.126 & 95.05 & 0.50 \\ 
  homogeneous & (40\%) & $\mc S$-score (BT-QT) & 0.03 & 0.059 & 94.65 & 0.23 & 0.16 & 0.126 & 94.80 & 0.49 \\ 
  across groups &  & Pocock & 0.05 & 0.069 & 94.80 & 0.27 & 7.57 & 0.204 & \textbf{77.30} & 0.51 \\
   & & Wang & 0.00 & 0.062 & \textbf{93.85} & 0.23 & 6.13 & 0.176 & \textbf{77.80} & 0.47 \\
   &  &  IPCW$^*$ &  0.27 &  0.068 &  94.80 &   0.27 &  7.12 &  0.198 &  \textbf{80.40} &  0.52 \\ 
    \addlinespace
   &  & $\mc S$-score (IF-Wald) & 0.56 & 0.059 & 94.90 & 0.24 & 0.22 & 0.126 & 94.90 & 0.50 \\ 
   & MAR & $\mc S$-score (BT-Wald) & 0.56 & 0.059 & 95.10 & 0.24 & 0.22 & 0.126 & 95.35 & 0.50 \\ 
   & (40\%) & $\mc S$-score (BT-QT) & 0.56 & 0.059 & 94.90 & 0.24 & 0.22 & 0.126 & 94.85 & 0.49 \\ 
   &  & Pocock & 8.66 & 0.107 & \textbf{73.65} & 0.25 & 11.20 & 0.265 & \textbf{55.85} & 0.49 \\ 
    & & Wang & 0.39 & 0.062 & 94.10 & 0.23 & 6.06 & 0.175 & \textbf{77.80} & 0.47 \\ 
    &  &  IPCW$^*$ &  8.30 &  0.105 &  \textbf{75.60} &  0.25 &  10.92 &  0.261 &  \textbf{59.60} &  0.50 \\ 
    \midrule 
   &  & $\mc S$-score (IF-Wald) & 0.14 & 0.060 & 95.45 & 0.24 & 0.13 & 0.125 & 94.90 & 0.50 \\ 
   & No & $\mc S$-score (BT-Wald) & 0.14 & 0.060 & 95.45 & 0.24 & 0.13 & 0.125 & 95.25 & 0.50 \\ 
   &  & $\mc S$-score (BT-QT) & 0.14 & 0.060 & 95.65 & 0.24 & 0.13 & 0.125 & 95.35 & 0.50 \\ 
   &  & Pocock & 7.95 & 0.102 & \textbf{76.35} & 0.25 & 5.92 & 0.180 & \textbf{84.50} & 0.53 \\ 
& & Wang & 20.79 & 0.222 & \textbf{17.15} & 0.29 & 7.89 & 0.218 & \textbf{83.65} & 0.55 \\  
&  &  IPCW$^*$ &  8.05 &   0.099 &  \textbf{77.20} &  0.25 &   6.34 &  0.182 &  \textbf{84.60} &  0.53 \\ 
    \addlinespace
  40\% censored $Y_1$,   &  & $\mc S$-score (IF-Wald) & 0.27 & 0.061 & 95.40 & 0.24 & 0.30 & 0.128 & 95.20 & 0.50 \\ 
  heterogeneous & MCAR & $\mc S$-score (BT-Wald) & 0.27 & 0.061 & 95.25 & 0.24 & 0.30 & 0.128 & 95.40 & 0.50 \\ 
   across groups & (40\%) & $\mc S$-score (BT-QT) & 0.27 & 0.061 & 94.85 & 0.24 & 0.30 & 0.128 & 94.40 & 0.50 \\ 
  (20\% vs. 60\%) &  & Pocock & 2.57 & 0.073 & \textbf{92.80} & 0.27 & 7.53 & 0.205 & \textbf{77.70} & 0.52 \\ 
  & & Wang & 20.61 & 0.220 & \textbf{18.25} & 0.29 & 8.13 & 0.224 & \textbf{80.60} & 0.56 \\
  &  &  IPCW$^*$ &  2.67 &   0.073 &  \textbf{93.00} &   0.27 &  7.76 &  0.210 & \textbf{78.20} &  0.52 \\ 
    \addlinespace
   &  & $\mc S$-score (IF-Wald) & 0.75 & 0.062 & 95.50 & 0.24 & 0.38 & 0.128 & 95.30 & 0.50 \\ 
   & MAR & $\mc S$-score (BT-Wald) & 0.75 & 0.062 & 95.55 & 0.25 & 0.38 & 0.128 & 95.30 & 0.50 \\ 
   & (40\%) & $\mc S$-score (BT-QT) & 0.75 & 0.062 & 95.00 & 0.24 & 0.38 & 0.128 & 94.45 & 0.50 \\ 
   &  & Pocock & 9.89 & 0.117 & \textbf{67.20} & 0.25 & 9.76 & 0.241 & \textbf{65.60} & 0.51 \\ 
   & & Wang & 20.84 & 0.223 & \textbf{16.85} & 0.29 & 8.18 & 0.225 & \textbf{80.40} & 0.56 \\
   &  &  IPCW$^*$ &  9.91 &  0.115 &  \textbf{64.80} &  0.25 &  9.90 &  0.243 &  \textbf{62.60} &  0.51 \\ 
  \bottomrule
  \end{tabular}
    \caption{Selected simulation results, with $(n_a,n_b)=(1000,1000)$. }
    \label{tab:simu-main}
    \begin{tablenotes}
	\scriptsize
	\item CP\%'s outside the interval [94, 96] are highlighted in bold. 
    \item 
    {$^*$The IPCW method is implemented in the \texttt{WINS} R package for the first endpoint $Y_1$ and excluding subjects with missing data in $Y_2$.  }
     \end{tablenotes}
\end{threeparttable}
\end{table}

We only present a subset of simulation results in Table \ref{tab:simu-main}, where we focus on the scenario of larger sample size ($n_a=n_b=1000$) with moderate average censoring and missing data rates (40\%). All the other results are shown in Online Supplemental Material (Web Appendix S.1).

In all three cases of censoring of $Y_1$, Pocock's method yields substantial biases and poor CP\%'s in the presence of missing $Y_2$ under both MCAR and MAR, except when $\WR=1$ under MCAR. With complete data on both $Y_1$ and $Y_2$ (the top case), our $\mc S$-score method and Pocock's method produce identical ARB\% and RMSE, with only minor differences in variance estimates, confirming their equivalence in point estimation under full data. When $Y_1$ is subject to censoring (either homogeneous or heterogeneous) across treatment groups, Pocock's method shows larger RMSEs for both $\WR=1$ and $2$, and poor CP\%'s when $\WR=2$. Moreover, when $\WR=2$, as long as there are censored data on $Y_1$ or missing data on $Y_2$, Pocock's method is biased, whereas our method maintains consistency and valid inference in all cases. 

In addition, regarding the three approaches for constructing 95\% CIs for our method (IF-Wald, BT-Wald and BT-QT), we observe negligible differences on their CP\% and CI width, supporting the validity of our proposed IF-based variance estimator and the utility of all three CI methods to quantify uncertainty. We observed that all CP\%'s from our $\mc S$-score methods are very close to the 95\% nominal level, lying within the [94, 96] range, except for three cases (96.40, 96.40, and 93.85, bolded in the table), which are also nearly 95\%. 

We observe that Wang's method can be biased and exhibits poor CP\%'s when $Y_1$ is censored, but remains overall robust when $Y_1$ is not censored, regardless of the missingness mechanism of $Y_2$. This is likely because the imputation model always assumes the censoring time follows an exponential distribution (but here the data were generated based on a Gamma distribution), while allowing imputation of missing data by treatment group (and is therefore robust to MAR in our simulations). Nevertheless, their method overall yields larger biases and RMSEs than our method. {Wang’s method also performs poorly compared to Pocock’s method in some scenarios, for example,} when the censoring is heterogeneous and $\WR=1$. While Wang's method may appear to perform better under heterogeneous censoring rates of 20\% (group $a$) versus 60\% (group $b$) when $\WR = 2$, by a simple experiment, we found that reversing the rates (60\% for group $a$ and 20\% for group $b$) leads to substantially larger bias and much poorer CP\%. A potential reason again is that imputing the wrong time to event may lead to imputing too many or too few events before $h$, biasing the win/loss counts and potentially leading to more biased estimator than Pocock's method. In the scenario of no censoring in $Y_1$ but with missing data in $Y_2$, Wang's method outperformed Pocock's method and yielded valid results. 

{Regarding the IPCW method implemented in the \texttt{WINS} package, we find that in the absence of censoring in $Y_1$ and missingness in $Y_2$, it yields identical ARB\% and RMSE to our $\mc S$-score and Pocock’s approaches, confirming their equivalence under complete data. Minor differences arise only in variance estimation, as reflected by the slightly higher CP\% values for IPCW. When $Y_2$ is fully observed and potential bias arises solely from censoring in $Y_1$, IPCW remains consistent under homogeneous censoring but becomes biased under heterogeneous censoring, whereas our method remains consistent. More generally, IPCW exhibits bias in the presence of both censoring in $Y_1$ and missingness in $Y_2$, potentially because IPCW method can only be applied to the subset of subjects without missing data in $Y_2$. When the true WR equals 1, IPCW performs well under MCAR and homogeneous censoring but is sensitive to heterogeneous censoring and missingness. When the true WR equals 2, either censoring or missingness leads to substantially inflated ARB\% and RMSE and to undercoverage. Overall, these results demonstrate the robustness and superior performance of our method across the simulation settings considered, relative to competing approaches. }

Moreover, although we do not display results from additional scenarios, such as smaller sample sizes ($n_a = n_b = 100$) or lower rates of censoring and missingness (20\%), these can be found in Online Supplemental Material (Web Appendix S.1). We briefly summarize the findings here. Overall, the patterns of bias in Pocock's method remain consistent with those shown in Table \ref{tab:simu-main}. Our method consistently has smaller biases and RMSEs. In smaller samples ($n_a = n_b = 100$), the three CI estimators (IF-Wald, BT-Wald and BT-QT) for our $\mc S$-score method occasionally yield coverage slightly outside [94, 96] (lowest 93.55, highest 96.45), likely due to {Monte Carlo variability, given that the $[94, 96]$ range is obtained from a normal approximation to the sampling distribution of the estimated CP\%}. Nevertheless, most CP\% values fall within [94, 96] and are close to the 95\% nominal level. The proposed IF-based variance estimator consistently achieves coverage in [93.65, 96.00], {with only two scenarios falling slightly below 94 (93.90 and 93.65),} confirming its validity for inference. Compared with the variance estimated through bootstrap approaches, it also offers greater computational efficiency by avoiding repeated resampling. 

{Lastly, we conduct an additional experiment to evaluate the small-sample performance of the proposed IF-based variance estimator under reduced sample sizes. Specifically, we generate data with $n_a = n_b = 50$ and $n_a = n_b = 25$ and assess the CP\% of the three CI estimation methods for the $\mc S$-score estimator considered in the main simulations. When $n_a = n_b = 50$, the CP\% values based on the IF-based variance estimator remain close to the nominal level, falling within the range of $[94, 96]$ overall. However, we find that when the sample size is reduced to $n_a = n_b = 25$, the IF-based variance estimator generally yields CP\% values below the lower bound of the nominal range, although all values remain above 91.80. Detailed results for $n_a = n_b = 25$ are provided in the Online Supplemental Material (Web Appendix S.1.2). This behavior is expected, as the IF-based variance estimator relies on large-sample theory and may be less accurate in very small samples, reflecting a natural trade-off between computational efficiency and finite-sample precision. In contrast, the bootstrap achieves CP\% close to the nominal level and is not computationally burdensome in small-sample settings. Taken together, these findings suggest a practical guideline: the IF-based variance estimator is well suited for moderate to large sample sizes because of its computational efficiency and reliable coverage, whereas bootstrap-based variance estimation is preferable in small-sample applications. Our empirical findings suggest that if a total sample size of $n<100$, bootstrap approach is recommended. }

\section{Real Data Analysis}\label{sec:data}

We apply our proposed method to two case studies. The first study, the HEART-FID trial \citep{mentz2023ferric}, is a double-blind, randomized clinical trial that evaluated the effect of ferric carboxymaltose against a placebo in patients with heart failure, reduced ejection fraction, and iron deficiency. The second, the ISCHEMIA trial \citep{maron2020initial}, randomized patients with moderate or severe ischemia to either an initial invasive strategy (including angiography and revascularization when feasible) plus medical therapy or an initial conservative strategy of medical therapy alone, with angiography reserved in case of medical failure. 

We apply our $\mc S$-score method with and without adjusting covariates $Y_2$ missing data  as shown in Sections \ref{subsec:propose-obs} and \ref{subsec:covar}. To adjust for missing data, we use a logistic regression model with some pre-specified covariates. We also evaluate and compare three variance estimation methods (IF-Wald, BT-Wald, and BT-QT), which we also used in our simulation studies. For the bootstrap-based methods, we use 1,000 bootstrap data replicates.

\subsection{Study I: HEART-FID trial}\label{subsec:heart-fid}

The HEART-FID trial \citep{mentz2023ferric} randomized and enrolled 3,065 participants, with 1,532 in the treatment group. The primary analysis included three hierarchical endpoints: time-to-death within one year, number of heart failure hospitalizations within 12 months and change in 6-minute walk distance from baseline to month 6. 

For illustration of our method, we chose two hierarchical endpoints: time-to-death within 12 months ($Y_1$) and 6-minute walk distance at month 12 ($Y_2$). Participants are administratively censored at month 12 ($h=364$ days); based on the observed maximum 6-minute walk distance at month 12, we set $\tau = 864$.  To impute missing  $Y_2$ values, we consider two baseline covariates used by \cite{mentz2023ferric}: the 6-minute walk distance at baseline and the participants' geographic region (North America, Asia Pacific, or Europe). 

\begin{table}[ht]
\begin{threeparttable}
    \small
    \begin{tabular}{clcccccccccccc}
    \toprule
  \multicolumn{2}{c}{Variable (with mean (SD) or $n$ (\%))}   &  \makecell[c]{Overall \\ $(n=3,065)$} & \makecell[c]{Control \\ $(n=1,533)$} & \makecell[c]{Treated \\ $(n=1,532)$} \\
         \midrule
    \multirow{5}{*}{Covariate}     & Baseline 6-min. walk distance  & 274.3 (109.5) & 274.7 (109.4) & 273.9 (109.7)\\
     & Region &   &  &   \\
       &  \qquad  North America & 1,442 (47.0\%) & 721 (47.0\%) & 721 (47.1\%) \\
       & \qquad Asia Pacific & 210 (6.9\%) & 105 (6.8\%) & 105 (6.9\%) \\
       & \qquad  Europe & 1,413 (46.1\%) & 707 (46.1\%) & 706 (46.1\%) \\
       \midrule
       \multirow{2}{*}{$Y_1$} & Death before  $h=364$ days & 289 (9.4\%) & 158 (10.3\%) & 131 (8.6\%) \\
      & \qquad  Right-censoring before $h$ & 80 (2.6\%) & 49 (3.2\%) & 31 (2.0\%) \\
      \midrule
      \addlinespace
    \multirow{5}{*}{$Y_2$}   & & \multicolumn{3}{c}{\bf Among participants with $Y_1>h$} \\
      \addlinespace
    & & \makecell[c]{Overall \\ $(n=2,696)$} & \makecell[c]{Control \\ $(n=1,326)$} & \makecell[c]{Treated \\ $(n=1,370)$} \\
      \cmidrule(lr){3-5}
       & 6-min. walk distance at month 12 & 290.7 (115.8) & 293.1 (120.1) & 288.5 (111.5)\\ 
      & \qquad  Missing $Y_2$ data & 419 (15.5\%) & 208 (15.7\%) & 211 (15.4\%) \\
      \bottomrule
    \end{tabular}
    \caption{Summary statistics of the HEART-FID trial data. }
    \label{tab:sumstat-hfid}
    \begin{tablenotes}\scriptsize
        \item 
        For $Y_2$, the summary statistics excluded the missing data. 
         SD: standard deviation; $n$ (\%): count and percentage. 
    \end{tablenotes}
    \end{threeparttable}
\end{table}
Overall, less than 3\% of the participants were censored before $h=364$ days, as shown in Table \ref{tab:sumstat-hfid}, and about 15\% of participants had missing $Y_2$ values. Nevertheless, those missing $Y_2$ data were mostly in North America (21\%) than Europe or Asia Pacific (both 11\%). Besides, participants with missing data had on average lower 6-minute walk distances at baseline. The rates of death before $h=364$ days and the mean $Y_2$ (without missing data) between the treatment groups were not statistically different  (p-value = 0.10 and 0.35, respectively). The logistic regression of the missingness indicator on treatment, the two covariates, and their interactions ($Z$, $\mb X$, and $Z\cdot\mb X$), showed no significant interaction (p-values $=0.36$ and $0.72$) between the treatment and the baseline covariates, but there were significant covariate main effects (p-values $<0.001$). 

{To assess the robustness of this missingness model to potential violations of the MAR assumption, we calculated the E-value \citep{vanderweele2017sensitivity}, which was 1.63, indicating that an unmeasured factor would need to be associated with both the probability of observing the second endpoint and the outcome by a risk ratio of at least 1.63, conditional on observed covariates, to fully explain away the observed association. } 

Table \ref{tab:dataI-res} presents the analysis results using our $\mc S$-score method, the covariate-adjusted $\mc S$-score method, and Pocock's algorithm. Accounting for the small amount of censored $Y_1$ and the missing $Y_2$ values, with or without covariate adjustment, yields WR estimates that are slightly smaller and have narrower 95\% CIs than those from Pocock's algorithm. Nonetheless, the overall conclusion remains the same. These WR estimates are also smaller than those reported in the primary analysis of the original study \citep{mentz2023ferric} because we only used two endpoints here rather than the original three hierarchical endpoints and we used the 6-minute walk distance at month 12 rather than at {month 6.} 

\begin{figure}[H]
    \centering
    \includegraphics[width=0.75\linewidth]{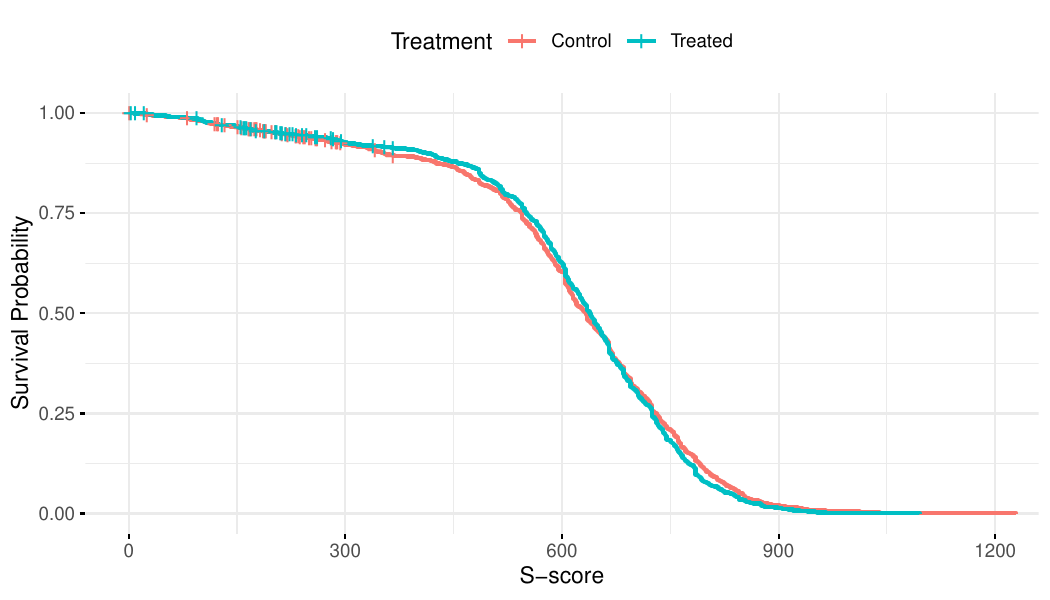}
    \caption{{Treatment-specific Kaplan-Meier curves of the $\mc S$-score for the HEART-FID study.}}
    \label{fig:km-FID}
\end{figure}

{To better understand the relationship between the estimated WR and the $\mc S$-score, we further plot treatment-specific Kaplan–Meier curves of the $\mc S$-score in Figure \ref{fig:km-FID}. As shown, the Kaplan–Meier curves for the two treatment groups are highly similar, which is consistent with the estimated WR being close to 1.  }

\begin{table}[H]
    \centering
    \begin{threeparttable} 
    \begin{tabular}{lcccccccccccccccc}
    \toprule
      Method & WR Estimate & 95\% CI & CI Width \\ 
      \midrule
       $\mc S$-score (IF-Wald) & 1.01 & (0.92, 1.10) & 0.18 \\ 
  $\mc S$-score (BT-Wald) & 1.01 & (0.92, 1.10) & 0.18 \\ 
  $\mc S$-score (BT-QT) & 1.01 & (0.93, 1.11) & 0.18 \\ 
  Covariate-adjusted $\mc S$-score (IF-Wald) & 1.02 & (0.94, 1.11) & 0.18 \\ 
  Covariate-adjusted $\mc S$-score (BT-Wald) & 1.02 & (0.93, 1.12) & 0.18 \\ 
  Covariate-adjusted $\mc S$-score (BT-QT) & 1.02 & (0.94, 1.12) & 0.18 \\ 
  Pocock & 1.04 & (0.96, 1.14) & 0.19 \\ 
       \bottomrule
    \end{tabular}
    \caption{HEART-FID trial: win ratio estimates for time-to-death by month 12 and 6-minute walk distance at month 12.}
    \label{tab:dataI-res}
    \begin{tablenotes}
	\scriptsize
	\item CI: confidence interval.
     \end{tablenotes}
\end{threeparttable}
\end{table}

This example illustrates a case where the true WR is not significantly different from  1, with low rates of $Y_1$ censoring  and moderate percentages of missing $Y_2$ data  in both groups. Thus, both the $\mc S$-score and Pocock's methods are likely consistent in this setting as shown in the simulation studies. However, we observe slightly narrow CIs with the $\mc S$-score method, indicating a potential gain in efficiency. This result aligns with our simulations, i.e., even when both methods are consistent, our $\mc S$-score method consistently yields more efficient estimates.

{To assess the sensitivity to potential violations of the non-informative censoring assumption, we conducted supplemental best- and worst-case analyses. These analyses impose extreme but plausible censoring mechanisms by assuming that all censored participants in the treatment group survived by the end of study and those censored participants in the control group died the next day (best-case analysis), with the assumptions reversed in the worst-case analysis. In this example, the best-case analysis yields a WR estimate of 1.02 (95\% CI: [0.93, 1.11]), whereas the worst-case analysis yields 1.01 (95\% CI: [0.92, 1.10]), compared with the primary analysis result of 1.01 (95\% CI: [0.92, 1.10]). These results show tiny variation across the best-, worst-, and primary analyses, indicating that the estimated WR is highly insensitive to extreme censoring assumptions in this setting and providing support for the plausibility of the non-informative censoring assumption for this example. }

Finally, to assess potential misspecification of the assumed MAR model, we conducted supplementary analyses for the covariate-adjusted $\mc S$-score estimator by varying the missingness model for $Y_2$ in Online Supplemental Material (Web Appendix S.2). Specifically, we considered several alternative missingness model specifications implemented in the \texttt{SuperLearner} R package \citep{van2007super}, including (i) a logistic regression model with all two-way interaction terms among covariates (\texttt{SL.glm.interaction}); (ii) a random forest learner (\texttt{SL.randomForest}); and (iii) an ensemble learner combining logistic regression without interaction terms (\texttt{SL.glm}), logistic regression with interaction terms (\texttt{SL.glm.interaction}), and random forest (\texttt{SL.randomForest}). The results indicate that the resulting WR estimates are largely consistent with those reported in Table \ref{tab:dataI-res}, with no meaningful changes across different missingness model specifications. These findings further support the robustness of the proposed estimator to reasonable misspecification of the missingness model.

\subsection{Study II: ISCHEMIA trial}

The ISCHEMIA trial randomized a total of 5,179 participants, with 2,588 assigned to the treatment group \citep{maron2020initial, spertus2020health}. In \cite{maron2020initial}, the primary endpoint was a composite of death from cardiovascular causes, myocardial infarction, or hospitalization for unstable angina, heart failure, or resuscitated cardiac arrest. The secondary endpoint combined death from cardiovascular causes and myocardial infarction. \cite{spertus2020health}, paid a special attention to  the analysis the Seattle Angina Questionnaire–Angina Frequency (SAQ7-AF) Score, a quality-of-life endpoint. 

For illustration, we chose two endpoints, the time to all-cause mortality  $Y_1$ within 4 years and the 4-year post-randomization SAQ7-AF score $Y_2$. 
We fixed $h$ to 1,460 days, at which points participants are administratively censored. Moreover, since $\max(Y_2)=100$, we set $\tau=100$. We also selected 5 covariates to adjust for missing  $Y_2$ data: age, sex, history of diabetes, ejection fraction, and estimated glomerular filtration rate. The summary statistics are given in Table \ref{tab:sumstat-isch} below.

\begin{table}[H]
\begin{threeparttable}
    \small
    \begin{tabular}{clcccccccccccc}
    \toprule
   \multicolumn{2}{c}{Variable (with mean (SD) or $n$ (\%))}  &  \makecell[c]{Overall \\ $(n=5,179)$} & \makecell[c]{Control \\ $(n=2,591)$} & \makecell[c]{Treated \\ $(n=2,588)$} \\
         \midrule
       & Age  & 64.3 (9.6) & 64.2 (9.7) & 64.4 (9.5)\\
       & Sex $=$ female  & 1,168 (22.6\%) & 562 (21.7\%) & 606 (23.4\%) \\
    Covariate   & History of diabetes & 2,164 (41.8\%) & 1,093 (42.2\%) & 1,071 (41.4\%) \\
       & Ejection fraction & 60.1 (8.2) & 59.9 (8.3) & 60.2 (8.1) \\
       & Glomerular filtration rate & 83.2 (22.6) & 83.1 (22.5) & 83.2 (22.8)  \\
       \midrule
       \multirow{2}{*}{$Y_1$}    & Death before $h=1,460$ days & 247 (4.8\%) & 123 (4.7\%) & 124 (4.8\%) \\
       &\qquad Right-censoring before $h$ & 3,264 (63.0\%) & 1,627 (62.8\%) & 1,637 (63.3\%) \\
      \midrule
      \addlinespace
       \multirow{5}{*}{$Y_2$}    & & \multicolumn{3}{c}{\bf Among participants with $Y_1>h$} \\
      \addlinespace
     & & \makecell[c]{Overall \\ $(n=1,688)$} & \makecell[c]{Control \\ $(n=841)$} & \makecell[c]{Treated \\ $(n=827)$} \\
      \cmidrule(lr){3-5}
       & SAQ7-AF score & 93.7 (13.4) & 92.8 (14.1) &  94.6 (12.5)\\ 
      & \qquad  Missing $Y_2$ data & 351 (21.0\%) & 176 (20.9\%) & 175 (21.2\%) \\
      \bottomrule
    \end{tabular}
    \caption{Summary statistics of the ISCHEMIA trial data. }
    \label{tab:sumstat-isch}
    \begin{tablenotes}\scriptsize
        \item SD: standard deviation; $n$ (\%): count and percentage. 
    \end{tablenotes}
    \end{threeparttable}
\end{table}

Overall, among the participants, about 5\% died  and 63\% were censored before  $h=1,460$ days, while 21\% had missing SAQ7-AF score. On average, there was not a statistical difference in covariates, all-cause death, the percentage of censored SAQ7-AF score before $h$ between the treatment groups. 
However, the groups differ significantly in their mean SAQ7-AF score (p-value = 0.01). The logistic regression model for the missing SAQ7-AF score indicator using baseline covariates, treatment assignment, and their interactions shows significant associations with age, history of diabetes, and glomerular filtration rate.

{Similar to the HEART-FID study, we assessed the E-value \citep{vanderweele2017sensitivity} for this missingness model, which was 1.98, indicating moderately strong robustness to potential violations of the MAR assumption. This result suggests that an unmeasured factor would need to be associated with both the probability of observing the second endpoint and the outcome by a risk ratio of at least 1.98, conditional on observed covariates, to fully explain away the observed association.} 

\begin{table}[H]
\begin{threeparttable}
    \centering
    \begin{tabular}{lccc}
    \toprule
      Method & Estimate & 95\% CI & CI Width \\ 
      \midrule
       $\mc S$-score (IF-Wald) & 1.26 & (1.04, 1.49) & 0.45 \\ 
  $\mc S$-score (BT-Wald) & 1.26 & (1.04, 1.48) & 0.44 \\ 
  $\mc S$-score (BT-QT) & 1.26 & (1.06, 1.49) & 0.42 \\ 
  Covariate-adjusted $\mc S$-score (IF-Wald) & 1.27 & (1.05, 1.49) & 0.44 \\ 
  Covariate-adjusted $\mc S$-score (BT-Wald) & 1.27 & (1.05, 1.49) & 0.44 \\ 
  Covariate-adjusted $\mc S$-score (BT-QT) & 1.27 & (1.08, 1.50) & 0.42 \\ 
  Pocock & 1.00 & (0.90, 1.12) & 0.22 \\ 
       \bottomrule
    \end{tabular}
    \caption{ISCHEMIA trial: win ratio estimates for time-to-death and SAQ7-AF score at month 48. }
    \label{tab:dataII-res}
     \begin{tablenotes}
	\scriptsize
	\item CI: confidence interval.
     \end{tablenotes}
\end{threeparttable}
\end{table}

\begin{figure}[H]
    \centering
    \includegraphics[width=0.75\linewidth]{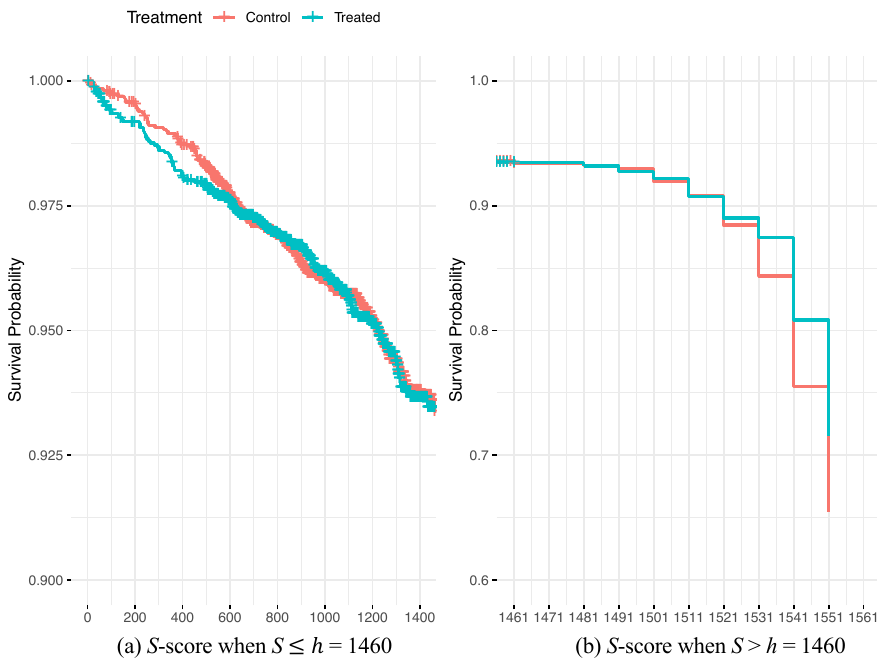}
    \caption{{Treatment-specific Kaplan-Meier curves of the $\mc S$-score for the ISCHEMIA study.}}
    \label{fig:km-ISC}
\end{figure}

Regarding the estimation results, our $\mc S$-score methods yield significantly different estimates compared to Pocock's algorithm (Table \ref{tab:dataII-res}). Both the unadjusted and covariate-adjusted $\mc S$-score methods produce similar point estimates, 1.26 and 1.27, that are statistically significant at the 0.05 level, while Pocock's method indicates a non significant result. The similarity between the adjusted and the non-covariate-adjusted results suggests that the adjustment model has a small impact on the estimation, likely because the rate of missing data does not differ significantly by treatment groups. 

{Similar to the HEART-FID study, we plot treatment-specific Kaplan–Meier curves of the $\mc S$-score in Figure \ref{fig:km-ISC}. To improve visualization of the survival probabilities, we split the overall curve into two panels, (a) and (b), using the threshold $h = 1460$ (the time horizon). This separation facilitates clearer inspection when $\mc S \leq h=1460$, where all survival probabilities exceed $0.9$. As shown in panel (a), the survival curves for the two treatment groups cross: the control group has higher survival at earlier times, while the treated group catches up and shows slightly higher survival after $\mc S \approx 800$. For $\mc S > 1000$, the two curves are approximately overlapping. As shown in panel (b), the treated group tends to have larger $\mc S$-scores, which is consistent with the QOL results in Table \ref{tab:sumstat-isch} (the treated group has a larger mean QOL value) and supports the plausibility of an estimated WR greater than $1$.
}

Our results are consistent with the finding of \cite{spertus2020health} where only the longitudinal QOLs were analyzed. Our results are also supported by our simulation findings, which suggest that Pocock's method may be biased when the true WR differs from 1 and missing or censored data are present regardless of whether such missingness or censoring is homogeneous or heterogeneous across treatment groups. Given the moderate to high rates of censoring and the percentage of the missing data in both groups in this example (see Table \ref{tab:sumstat-isch}), our simulations indicate that the $\mc S$-score method provides more reliable estimates than Pocock's algorithm.

{To assess sensitivity to potential violations of the non-informative censoring assumption, we conducted best- and worst-case analyses as described in Section \ref{subsec:heart-fid}. The best-case analysis yielded a WR estimate of 1.68 (95\% CI: [1.39, 1.98]), whereas the Worst analysis yielded 0.91 (95\% CI: [0.75, 1.07]), compared with the primary analysis result of 1.26 (95\% CI: [1.04, 1.49]). Confidence intervals were constructed using the proposed IF-based variance estimator. This large range of the estimated effect is reflective of a large percentage of censoring due to  these extreme scenarios. Additional information on the censoring mechanism is needed to narrow the range of the estimated effect. Therefore, the investigators should interpret the results with caution by  commenting on the plausibility of the non-informative censoring assumption.}

{Finally, similar to the HEART-FID analysis, we conducted supplementary analyses for the covariate-adjusted $\mc S$-score estimator by varying the missingness model for $Y_2$ in Online Supplemental Material (Web Appendix S.2). The alternative missingness model specifications considered were the same as those used in the HEART-FID analysis. The resulting WR estimates are very close to those reported in Table~\ref{tab:dataII-res}, indicating no meaningful differences across models. These findings further support the robustness of the proposed estimator to reasonable misspecification of the missingness model. }

\section{Concluding Remarks}\label{sec:concludes}

In this paper, we address a critical gap in estimating the WR for clinical trials involving two hierarchical endpoints, a time-to-terminal event subject to right-censoring and a non-survival outcome with missing data. By mapping the available information from the two endpoints to one endpoint, the $\mc S$-score, it enables us to construct simple NPMLE estimator for WR under non-informative censoring and MAR assumptions. This contrasts with the  counting-based approach used by Pocock's method that may discard observed data or valuable information to declared a pair as a win, loss, or tie, potentially leading to biased or less efficient estimates.

Through extensive simulations, we demonstrate that our method produces more consistent and efficient WR estimates, particularly in scenarios where the treatment effect is {WR $>1$}, the non-survival endpoint has sizable {percentages} of missing values (under both MCAR and MAR), or the rates of censoring on survival endpoint are not negligible. The simulation results also highlight the efficiency loss from Pocock's method in many such settings. Comparing our method to the imputation approach of \cite{wang2025restricted}, our method offers efficient computation time and greater robustness to model mis-specification. 

Furthermore, the proposed asymptotic variance estimator in Section \ref{subsec:varest} is consistent, enabling valid uncertainty quantification for the WR estimator. {However, in small-sample scenarios (such as the simulation with $n_a = n_b = 25$), it may exhibit slight undercoverage; therefore, we recommend using bootstrap-based variance estimation for total sample size less than 100.} Overall, our approach offers a practical and robust solution for analyzing two hierarchical endpoints without discarding any observed data, ultimately enabling more reliable estimation and inference on treatment effect via the WR.

{In addition, we applied our method to two real clinical trials, HEART-FID and ISCHEMIA, in Section \ref{sec:data}. These two studies illustrate complementary settings, covering scenarios with small versus substantial censoring and covariate-dependent missingness. Consistent with our simulation findings, our method potentially mitigates bias from Pocock’s counting approach. }

We acknowledge several limitations of our study that require future research. First, the proposed $\mc S$-score method is specifically tailored to the two-endpoint setting considered in this work and currently lacks broader generalizability. Extending or adapting the approach to accommodate more complex hierarchical structures, such as multi-categorical endpoints \citep{li2024elusiveness} or scenarios involving more than two endpoints, is non-trivial. In future work, we plan to explore distribution-based approach to estimate WR within a more flexible framework capable of handling a wider range of hierarchical endpoint configurations. Second, we have not extended covariate adjustments that include modeling the censoring distribution of $Y_1$ by $\mb X$ similar to the results of Section \ref{subsec:covar}. This omission is intentional;  modeling the censoring probability at each time point as a function of covariates (possibly time-varying) is more challenging and requires estimating a number of additional nuisance parameters. {Third, we acknowledge that violations of non-informative and/or MAR assumptions can compromise identification and introduce bias in the S-score estimator. More formal sensitivity analyses on these assumptions \citep{robins2000sensitivity} represent important extensions of the current framework and will be explored in future work. In addition, we aim to develop doubly robust (or even multiply robust) estimators \citep{robins1994estimation, bai2013doubly} that combine models for the outcome and the missingness mechanism, thereby further improving robustness to model misspecification. }

Finally, several potential extensions are worth pursuing for future research. The general strategy underlying our method might be adapted for WR analyses in observational studies (without randomization), multi-center clinical trials, and targeted subgroup analysis \citep{zhuang2025assessment, liu2025targeted, liu2024multi}. It is also interesting to consider extending nonparametric methods to Wang's imputation-then-counting approach \citep{wang2025restricted} in order to guard against misspecification in the imputation model by using, for instance, sample-splitting with machine learning tools \citep{van2007super, chernozhukov2018double, wang2025rate}.  

\subsection*{Data Availability Statement}

Data used in this paper are not publicly available but can be requested from the trials' sponsors.

\subsection*{Software}

A user-friendly R package to implement the proposed methodology is available at the author's Github page \url{https://github.com/yiliu1998/WinRS}.

\subsection*{Acknowledgment}

Yi Liu was supported by the National Heart, Lung, and Blood Institute (NHLBI) of the National Institutes of Health (NIH) under Award Number T32HL079896. The content of this paper is solely the responsibility of the authors and does not necessarily represent the official views of the NIH. The Duke Clinical Research Institute Biostatistics \& Data Science Research Fund also supported this work.

\bibliography{refs}

@book{Buyse2025GPC,
  title     = {Handbook of Generalized Pairwise Comparisons: Methods for Patient-Centric Analysis},
  editor    = {Marc Buyse and Johan Verbeeck and Everardo D. Saad and Mickaël De Backer and Vaiva Deltuvaite-Thomas and Geert Molenberghs},
  year      = {2025},
  publisher = {Chapman and Hall/CRC},
  address   = {New York},
  isbn      = {9781003390855},
  doi       = {10.1201/9781003390855}
}

@inproceedings{robins2000sensitivity,
	address = {New York, NY},
	author = {Robins, James M. and Rotnitzky, Andrea and Scharfstein, Daniel O.},
	booktitle = {Statistical Models in Epidemiology, the Environment, and Clinical Trials},
	editor = {Halloran, M. Elizabeth and Berry, Donald},
	isbn = {978-1-4612-1284-3},
	pages = {1--94},
	publisher = {Springer New York},
	title = {Sensitivity Analysis for Selection bias and unmeasured Confounding in missing Data and Causal inference models},
	year = {2000}}

@article{vanderweele2017sensitivity,
  title={Sensitivity analysis in observational research: introducing the E-value},
  author={VanderWeele, Tyler J and Ding, Peng},
  journal={Annals of internal medicine},
  volume={167},
  number={4},
  pages={268--274},
  year={2017},
  publisher={American College of Physicians}
}

@article{matsouaka2022robust,
  title={Robust statistical inference for matched win statistics},
  author={Matsouaka, Roland A},
  journal={Statistical Methods in Medical Research},
  volume={31},
  number={8},
  pages={1423--1438},
  year={2022},
  publisher={SAGE Publications Sage UK: London, England}
}

@article{liu2025targeted,
  title={Targeted Data Fusion for Causal Survival Analysis Under Distribution Shift},
  author={Liu, Yi and Levis, Alexander W and Zhu, Ke and Yang, Shu and Gilbert, Peter B and Han, Larry},
  journal={arXiv preprint arXiv:2501.18798},
  year={2025}
}

@article{dong2020inverse,
  title={The inverse-probability-of-censoring weighting (IPCW) adjusted win ratio statistic: an unbiased estimator in the presence of independent censoring},
  author={Dong, Gaohong and Mao, Lu and Huang, Bo and Gamalo-Siebers, Margaret and Wang, Jiuzhou and Yu, GuangLei and Hoaglin, David C},
  journal={Journal of Biopharmaceutical Statistics},
  volume={30},
  number={5},
  pages={882--899},
  year={2020},
  publisher={Taylor \& Francis}
}

@article{dong2021adjusting,
  title={Adjusting win statistics for dependent censoring},
  author={Dong, Gaohong and Huang, Bo and Wang, Duolao and Verbeeck, Johan and Wang, Jiuzhou and Hoaglin, David C},
  journal={Pharmaceutical Statistics},
  volume={20},
  number={3},
  pages={440--450},
  year={2021},
  publisher={Wiley Online Library}
}

@article{monzo2024use,
  title={Use of the win ratio analysis in critical care trials},
  author={Monzo, Luca and Levy, Bruno and Duarte, Kevin and Baudry, Guillaume and Combes, Alain and Ouattara, Alexandre and Delmas, Cl{\'e}ment and Kimmoun, Antoine and Girerd, Nicolas},
  journal={American Journal of Respiratory and Critical Care Medicine},
  volume={209},
  number={7},
  pages={798--804},
  year={2024},
  publisher={American Thoracic Society}
}

@article{duarte2022winratio,
  title={WinRatio: Win Ratio for Prioritized Outcomes and 95\% Confidence Interval},
  author={Duarte, K and Ferreira, JP},
  journal={The Comprehensive R Archive Network (CRAN)},
  year={2022},
  url={https://cran.r-project.org/web/packages/WinRatio/}
}

@article{gao2024does,
  title={When does adjusting covariate under randomization help? A comparative study on current practices},
  author={Gao, Ying and Liu, Yi and Matsouaka, Roland},
  journal={BMC Medical Research Methodology},
  volume={24},
  number={1},
  pages={250},
  year={2024},
  publisher={Springer}
}

@article{wang2025rate,
  title={Rate doubly robust estimation for weighted average treatment effects},
  author={Wang, Yiming and Liu, Yi and Yang, Shu},
  journal={Journal of Causal Inference},
  volume={13},
  number={1},
  pages={20240073},
  year={2025},
  publisher={De Gruyter}
}

@article{van2007super,
  title={Super learner},
  author={Van der Laan, Mark J and Polley, Eric C and Hubbard, Alan E},
  journal={Statistical applications in genetics and molecular biology},
  volume={6},
  number={1},
  year={2007},
  pages={1--23},
  publisher={De Gruyter}
}

@article{spertus2020health,
  title={Health-status outcomes with invasive or conservative care in coronary disease},
  author={Spertus, John A and Jones, Philip G and Maron, David J and O’Brien, Sean M and Reynolds, Harmony R and Rosenberg, Yves and Stone, Gregg W and Harrell Jr, Frank E and Boden, William E and Weintraub, William S and others},
  journal={New England Journal of Medicine},
  volume={382},
  number={15},
  pages={1408--1419},
  year={2020},
  publisher={Mass Medical Soc}
}

@article{mentz2023ferric,
  title={Ferric carboxymaltose in heart failure with iron deficiency},
  author={Mentz, Robert J and Garg, Jyotsna and Rockhold, Frank W and Butler, Javed and De Pasquale, Carmine G and Ezekowitz, Justin A and Lewis, Gregory D and O’Meara, Eileen and Ponikowski, Piotr and Troughton, Richard W and others},
  journal={New England Journal of Medicine},
  volume={389},
  number={11},
  pages={975--986},
  year={2023},
  publisher={Mass Medical Soc}
}

@article{kahan2024estimands,
  title={The estimands framework: a primer on the {ICH E9 (R1)} addendum},
  author={Kahan, Brennan C and Hindley, Joanna and Edwards, Mark and Cro, Suzie and Morris, Tim P},
  journal={BMJ},
  volume={384},
  year={2024},
  publisher={British Medical Journal Publishing Group}
}

@article{dong2020winimpact,
  title={The win ratio: Impact of censoring and follow-up time and use with nonproportional hazards},
  author={Dong, Gaohong and Huang, Bo and Chang, Yu-Wei and Seifu, Yodit and Song, James and Hoaglin, David C},
  journal={Pharmaceutical Statistics},
  volume={19},
  number={3},
  pages={168--177},
  year={2020},
  publisher={Wiley Online Library}
}

@article{maron2020initial,
  title={Initial invasive or conservative strategy for stable coronary disease},
  author={Maron, David J and Hochman, Judith S and Reynolds, Harmony R and Bangalore, Sripal and O’Brien, Sean M and Boden, William E and Chaitman, Bernard R and Senior, Roxy and L{\'o}pez-Send{\'o}n, Jose and Alexander, Karen P and others},
  journal={New England Journal of Medicine},
  volume={382},
  number={15},
  pages={1395--1407},
  year={2020},
  publisher={Mass Medical Soc}
}

@article{wang2025restricted,
  title={Restricted Time Win Ratio: From Estimands to Estimation},
  author={Wang, Tuo and Li, Ying and Qu, Yongming},
  journal={Statistics in Biopharmaceutical Research},
  volume={17},
  number={1},
  pages={136--148},
  year={2025},
  publisher={Taylor \& Francis}
}

@article{cui2025wins,
  title={WINS: The R WINS Package},
  author={Cui, Ying and Huang, Bo},
  year={2025},
  journal={The Comprehensive R Archive Network (CRAN)},
  url={https://cran.r-project.org/web/packages/WINS/}
}

@article{ferreira2020use,
  title={Use of the win ratio in cardiovascular trials},
  author={Ferreira, Jo{\~a}o Pedro and Jhund, Pardeep S and Duarte, K{\'e}vin and Claggett, Brian L and Solomon, Scott D and Pocock, Stuart and Petrie, Mark C and Zannad, Faiez and McMurray, John JV},
  journal={Heart failure},
  volume={8},
  number={6},
  pages={441--450},
  year={2020},
  publisher={American College of Cardiology Foundation Washington DC}
}

@article{krittayaphong2024components,
  title={Components of the Atrial fibrillation Better Care pathway for holistic care of patients with atrial fibrillation: a win ratio analysis from the COOL-AF registry},
  author={Krittayaphong, Rungroj and Treewaree, Sukrit and Lip, Gregory YH},
  journal={Europace},
  volume={26},
  number={9},
  pages={euae237},
  year={2024},
  publisher={Oxford University Press UK}
}

@article{abdalla2016win,
  title={The win ratio approach to analyzing composite outcomes: an application to the EVOLVE trial},
  author={Abdalla, Safa and Montez-Rath, Maria E and Parfrey, Patrick S and Chertow, Glenn M},
  journal={Contemporary Clinical Trials},
  volume={48},
  pages={119--124},
  year={2016},
  publisher={Elsevier}
}

@article{liu2025coadvise,
  title={COADVISE: Covariate Adjustment with Variable Selection and Missing Data Imputation in Randomized Controlled Trials},
  author={Liu, Yi and Zhu, Ke and Han, Larry and Yang, Shu},
  journal={arXiv preprint arXiv:2501.08945},
  year={2025}
}

@article{li2024elusiveness,
  title={The Elusiveness of the Win Ratio Parameter in the Presence of Missing Data},
  author={Li, Heng and Chen, Wei-Chen and Lu, Nelson and Tang, Rong and Zhao, Yu},
  journal={Therapeutic Innovation \& Regulatory Science},
  volume={58},
  number={3},
  pages={431--432},
  year={2024},
  publisher={Springer}
}

@article{redfors2020win,
  title={The win ratio approach for composite endpoints: practical guidance based on previous experience},
  author={Redfors, Bj{\"o}rn and Gregson, John and Crowley, Aaron and McAndrew, Thomas and Ben-Yehuda, Ori and Stone, Gregg W and Pocock, Stuart J},
  journal={European Heart Journal},
  volume={41},
  number={46},
  pages={4391--4399},
  year={2020},
  publisher={Oxford University Press}
}

@article{bebu2016large,
  title={Large sample inference for a win ratio analysis of a composite outcome based on prioritized components},
  author={Bebu, Ionut and Lachin, John M},
  journal={Biostatistics},
  volume={17},
  number={1},
  pages={178--187},
  year={2016},
  publisher={Oxford University Press}
}

@article{anker2009ferric,
  title={Ferric carboxymaltose in patients with heart failure and iron deficiency},
  author={Anker, Stefan D and Comin Colet, Josep and Filippatos, Gerasimos and Willenheimer, Ronnie and Dickstein, Kenneth and Drexler, Helmut and L{\"u}scher, Thomas F and Bart, Boris and Banasiak, Waldemar and Niegowska, Joanna and others},
  journal={New England Journal of Medicine},
  volume={361},
  number={25},
  pages={2436--2448},
  year={2009},
  publisher={Mass Medical Soc}
}

@article{barnhart2025trial,
  title={Trial design with win statistics for multiple time-to-event endpoints with hierarchy},
  author={Barnhart, Huiman X and Lokhnygina, Yuliya and Matsouaka, Roland A and Rockhold, Frank W},
  journal={Statistics in Biopharmaceutical Research},
  volume={17},
  number={2},
  pages={197--210},
  year={2025},
  publisher={Taylor \& Francis}
}

@article{zhuang2025assessment,
  title={Assessment of treatment effect heterogeneity for multiregional randomized clinical trials},
  author={Zhuang, Haotian and Wang, Xiaofei and George, Stephen L},
  journal={Statistics in Biopharmaceutical Research},
  volume={17},
  number={3},
  pages={315--322},
  year={2025},
  publisher={Taylor \& Francis}
}

@article{fine1999proportional,
  title={A proportional hazards model for the subdistribution of a competing risk},
  author={Fine, Jason P and Gray, Robert J},
  journal={Journal of the American statistical association},
  volume={94},
  number={446},
  pages={496--509},
  year={1999},
  publisher={Taylor \& Francis}
}

@article{dong2023win,
  title={Win statistics (win ratio, win odds, and net benefit) can complement one another to show the strength of the treatment effect on time-to-event outcomes},
  author={Dong, Gaohong and Huang, Bo and Verbeeck, Johan and Cui, Ying and Song, James and Gamalo-Siebers, Margaret and Wang, Duolao and Hoaglin, David C and Seifu, Yodit and M{\"u}tze, Tobias and others},
  journal={Pharmaceutical Statistics},
  volume={22},
  number={1},
  pages={20--33},
  year={2023},
  publisher={Wiley Online Library}
}

@article{hougaard1999multi,
  title={Multi-state models: a review},
  author={Hougaard, Philip},
  journal={Lifetime Data Analysis},
  volume={5},
  pages={239--264},
  year={1999},
  publisher={Springer}
}

@article{putter2007tutorial,
  title={Tutorial in biostatistics: competing risks and multi-state models},
  author={Putter, Hein and Fiocco, Marta and Geskus, Ronald B},
  journal={Statistics in Medicine},
  volume={26},
  number={11},
  pages={2389--2430},
  year={2007},
  publisher={Wiley Online Library}
}

@article{gooley1999estimation,
  title={Estimation of failure probabilities in the presence of competing risks: new representations of old estimators},
  author={Gooley, Ted A and Leisenring, Wendy and Crowley, John and Storer, Barry E},
  journal={Statistics in Medicine},
  volume={18},
  number={6},
  pages={695--706},
  year={1999},
  publisher={Wiley Online Library}
}

@article{bai2013doubly,
  title={Doubly-robust estimators of treatment-specific survival distributions in observational studies with stratified sampling},
  author={Bai, Xiaofei and Tsiatis, Anastasios A and O'Brien, Sean M},
  journal={Biometrics},
  volume={69},
  number={4},
  pages={830--839},
  year={2013},
  publisher={Oxford University Press}
}

@article{kaplan1958nonparametric,
  title={Nonparametric estimation from incomplete observations},
  author={Kaplan, Edward L and Meier, Paul},
  journal={Journal of the American Statistical Association},
  volume={53},
  number={282},
  pages={457--481},
  year={1958},
  publisher={Taylor \& Francis}
}

@article{cox1972regression,
  title={Regression models and life-tables},
  author={Cox, David R},
  journal={Journal of the Royal Statistical Society: Series B (Methodological)},
  volume={34},
  number={2},
  pages={187--202},
  year={1972},
  publisher={Wiley Online Library}
}

@article{finkelstein1999combining,
  title={Combining mortality and longitudinal measures in clinical trials},
  author={Finkelstein, Dianne M and Schoenfeld, David A},
  journal={Statistics in medicine},
  volume={18},
  number={11},
  pages={1341--1354},
  year={1999},
  publisher={Wiley Online Library}
}

@article{pocock2012win,
  title={The win ratio: a new approach to the analysis of composite endpoints in clinical trials based on clinical priorities},
  author={Pocock, Stuart J and Ariti, Cono A and Collier, Timothy J and Wang, Duolao},
  journal={European Heart Journal},
  volume={33},
  number={2},
  pages={176--182},
  year={2012},
  publisher={Oxford University Press}
}

@article{rubin1976inference,
  title={Inference and missing data},
  author={Rubin, Donald B},
  journal={Biometrika},
  volume={63},
  number={3},
  pages={581--592},
  year={1976},
  publisher={Oxford University Press}
}

@book{van2000asymptotic,
  title={Asymptotic Statistics},
  author={Van der Vaart, Aad W},
  volume={3},
  year={2000},
  publisher={Cambridge university press}
}

@book{shao2012jackknife,
  title={The jackknife and bootstrap},
  author={Shao, Jun and Tu, Dongsheng},
  year={2012},
  publisher={Springer Science \& Business Media}
}

@book{efron1994introduction,
	title={An introduction to the bootstrap},
	author={Efron, Bradley and Tibshirani, Robert J},
	year={1994},
	publisher={CRC press}
}

@article{robins1994estimation,
	title={Estimation of regression coefficients when some regressors are not always observed},
	author={Robins, James M and Rotnitzky, Andrea and Zhao, Lue Ping},
	journal={Journal of the American statistical Association},
	volume={89},
	number={427},
	pages={846--866},
	year={1994},
	publisher={Taylor \& Francis}
}

@article{stefanski2002calculus,
	title={The calculus of M-estimation},
	author={Stefanski, Leonard A and Boos, Dennis D},
	journal={The American Statistician},
	volume={56},
	number={1},
	pages={29--38},
	year={2002},
	publisher={Taylor \& Francis}
}

@book{tsiatis2007semiparametric,
	title={Semiparametric theory and missing data},
	author={Tsiatis, Anastasios},
	year={2007},
	publisher={Springer Science \& Business Media}
}

@article{liu2024multi,
  title={Multi-source conformal inference under distribution shift},
  author={Liu, Yi and Levis, Alexander W and Normand, Sharon-Lise and Han, Larry},
  journal={Proceedings of Machine Learning Research},
  volume={235},
  pages={31344--31382},
  year={2024}
}

@article{mao2024defining,
  title={Defining estimand for the win ratio: Separate the true effect from censoring},
  author={Mao, Lu},
  journal={Clinical Trials},
  volume={21},
  number={5},
  pages={584--594},
  year={2024},
  publisher={SAGE Publications Sage UK: London, England}
}

@article{chernozhukov2018double,
  title={Double/debiased machine learning for treatment and structural parameters},
  author={Chernozhukov, Victor and Chetverikov, Denis and Demirer, Mert and Duflo, Esther and Hansen, Christian and Newey, Whitney and Robins, James},
  year={2018},
    journal={The Econometrics Journal},
  publisher={Oxford University Press Oxford, UK}
}
\newpage

\appendix\label{Appendix}
\counterwithin{equation}{section}

\section{Appendix: Technical Proofs}

\subsection{Proof of Equation \eqref{thm:equi}}\label{subapp:proof-equi}

First, let us show that $Y_{1b}<Y_{1a}$ or $Y_{1b}=Y_{1a}, Y_{2b}<Y_{2a}$ implies $\mc S_b<\mc S_a$ with probability 1. 

If $Y_{1b}<Y_{1a}\leq h$, then $\mc S_b=Y_{1b}<Y_{1a}=\mc S_a$. If $Y_{1b}<Y_{1a}=h+1$, we have $\mc S_b=Y_{1b}<Y_{1a} + Y_{2a}=\mc S_a$, since $Y_{2a}>0$. Now, if $Y_{1b}=Y_{1a}=h+1$ and $Y_{2b}<Y_{2a}$, then $\mc S_b=h+Y_{2b}<h+Y_{2a}=\mc S_a$. Lastly, the event $\{Y_{1b}=Y_{1a}\leq h, Y_{2b}<Y_{2a}\}$ has a zero probability given $Y_1$ is a terminal event and we do not consider the hypothetical existence of $Y_2$ before time $h$, thus we could omit to discuss this case. 

Now, let show that $\mc S_b<\mc S_a$ implies $Y_{1b}<Y_{1a}$ or $Y_{1b}=Y_{1a}, Y_{2b}<Y_{2a}$. \\
First, if $\mc S_a\geq h+1$, then the only possibility is that $Y_{1a}=h+1$ and there exists $Y_{2a}\geq 0$ such that $\mc S_a=h+1+Y_{2a}$. In this case, if $\mc S_b\geq h+1$ as well, there exists $Y_{2b}\geq 0$ such that $h+Y_{2b}=\mc S_b<\mc S_a=h+Y_{2a}$, which shows that $Y_{1b}=Y_{1a}=h+1$ as well as $Y_{2b}<Y_{2a}$. However, if $\mc S_b\leq h$, then $Y_{1b}\leq h$, implying that $Y_{1b}<Y_{1a}=h+1$. Second, if $\mc S_a\leq h$, then $\mc S_b\leq h$. Since, in this case, we can only have that $\mc S_a=Y_{1a}$ and $\mc S_b=Y_{1b}$. \\Therefore, $Y_{1b}<Y_{1a}$. 

The proof is completed. 

\subsection{Consistency of the estimator $\WRhat(\mc O^{\mc S})$ in \eqref{eq:WR-npmle-obs}}\label{subapp:proof-consistency}

We adopt all notation in Section \ref{sec:method}. 
To prove that the proposed NPMLE \eqref{eq:WR-npmle-obs} is consistent for $\WR$, we first note that for $z\in\{a,b\}$ and under Assumptions \ref{asp:censor} and \ref{asp:MAR}, without proof,
\begin{align}\label{eq:Fz-converge}
    \sup_{s\in[0,h+\tau+1]}\left\vert\widehat F_z(s)-F_z(s)\right\vert\xrightarrow{p}0, \quad \text{as } n\to\infty.
\end{align}

We next establish the consistency of the ``plug‐in estimator'' \eqref{eq:WR-npmle} in Section \ref{subsec:propose-obs} (by plugging-in the above Kaplan-Meier estimator for the CDF $\widehat F_z$ to corresponding counterparts). We first consider the numerator of \eqref{eq:WR-npmle}. Define the target functional by
$$
\phi(F_a,F_b)=\int_0^{h+\tau+1}\{1-F_a(s)\}\, dF_b(s),
$$
and note that the plug‐in estimator is given by
$$
\widehat \phi = \phi(\widehat F_a,\widehat F_b)=\int_0^{h+\tau+1}\{1-\widehat F_a(s)\}\, d\widehat F_b(s).
$$
Our goal is to show $\widehat \phi\xrightarrow{p} \phi(F_a,F_b).$ To that end, consider the difference $\left|\widehat\phi-\phi(F_a,F_b)\right|.$ By adding and subtracting an intermediate term, we obtain
\begin{align*}
&\Biggl|\int_0^{h+\tau+1}\{1-\widehat F_a(s)\}\, d\widehat F_b(s)-\int_0^{h+\tau+1}\{1-F_a(s)\}\, dF_b(s)\Biggr|\\
&\leq \Biggl|\int_0^{h+\tau+1}\{1-\widehat F_a(s)\}\, d\widehat F_b(s)-\int_0^{h+\tau+1}\{1-\widehat F_a(s)\}\, dF_b(s)\Biggr|\\
& \quad + \Biggl|\int_0^{h+\tau+1}\{1-\widehat F_a(s)\}\, dF_b(s)-\int_0^{h+\tau+1}\{1-F_a(s)\}\, dF_b(s)\Biggr|.
\end{align*}
For the first term, observe that since $0\leq 1-\widehat F_a(s)\leq 1$ for all $t$, it follows that
$$
\Biggl|\int_0^{h+\tau+1}\{1-\widehat F_a(s)\}\, d\widehat F_b(s)-\int_0^{h+\tau+1}\{1-\widehat F_a(s)\}\, dF_b(s)\Biggr|
\leq \sup_{s\in[0,h+\tau+1]}\left|\widehat F_b(s)-F_b(s)\right| 
$$
By the uniform consistency of $\widehat F_b(s)$, the right-hand side converges in probability to zero.

For the second term, we have
$$
\Biggl|\int_0^{h+\tau+1}\{1-\widehat F_a(s)\}\, dF_b(s)-\int_0^{h+\tau+1}\{1-F_a(s)\}\, dF_b(s)\Biggr|
=\left|\int_0^{h+\tau+1}\bigl(\widehat F_a(s)-F_a(s)\bigr)\, dF_b(s)\right|.
$$
Since $dF_b(s)$ is a finite measure on $[0,h+\tau+1]$, it follows that
$$
\left|\int_0^{h+\tau+1}\bigl(\widehat F_a(s)-F_a(s)\bigr)\, dF_b(s)\right|
\leq \sup_{s\in[0,h+\tau+1]}\left|\widehat F_a(s)-F_a(s)\right|\Bigl(F_b(h+\tau+1)-F_b(0)\Bigr).
$$
By the uniform consistency of $\widehat F_a(s)$ and the finiteness of $F_b(h+\tau+1)-F_b(0)$, this term also converges in probability to zero.

Combining these results, we conclude that
$$
\left|\widehat \phi-\phi(F_a,F_b)\right|\xrightarrow{p} 0,
$$
which establishes that $\displaystyle\int_0^{h+\tau+1}\{1-\widehat F_a(s)\}\, d\widehat F_b(s)\xrightarrow{p} P(\mc S_a>\mc S_b)$. Similarly, we can show $\displaystyle\int_0^{h+\tau+1}\{1-\widehat F_b(s)\}\, d\widehat F_a(s)\xrightarrow{p} P(\mc S_a<\mc S_b)$. Hence, by Slutsky's theorem, $\WRhat\xrightarrow{p}\WR$ as $n\to\infty$. 

Finally, if the survival time is known to be discrete where both $F_a(s)$ and $F_b(s)$ functions are assumed as the CDFs of some probability mass functions (pmfs), we can replace the integral sign $\displaystyle\int$ by $\displaystyle\sum$ in above proof. Specifically, without loss of generality, assuming the time take values from $\{0,1,\dots,h+\tau+1\}$, we could re-write the above plug-in estimator as
\begin{align*}
    \widehat\phi=\phi(\widehat F_a,\widehat F_b)=\sum_{s=0}^{h+\tau+1}\{1-\widehat F_a(s)\}\{\widehat F_b(s)-\widehat F_b(s-1)\}.
\end{align*}
The remainder of the proof is similar. 

\subsection{Asymptotic variance estimation based on derived influence function}\label{subapp:varest}

Recall that the WR defined by the $\mc S$-score is expressed as
\begin{align*} 
\WR = \frac{P(\mc S_a > \mc S_b)}{P(\mc S_a < \mc S_b)} = \frac{\displaystyle\int \bar{F}_a(s)  dF_b(s)}{\displaystyle\int \bar{F}_b(s)  dF_a(s)}. \end{align*} For simplicity, we denote
\begin{align} \label{eq:ND}
N = \int \bar{F}_a(s)  dF_b(s), \quad D = \int \bar{F}_b(s)  dF_a(s), \end{align} so that the WR is given by $\WR = N/D$. 

We derive the variance of the WR estimator under the assumption that the $\mc S$-score takes values on a finite support. This is not restrictive in practice, as observed outcomes necessarily have finite precision and thus lie in a finite, albeit potentially very large, set. Treating $\mc S$ as a discrete random variable on a finite grid avoids measure-theoretic complications while accurately reflecting the observable nature of trial data. This formulation provides a rigorous foundation for applying M-estimation theory to derive the asymptotic variance of the estimator.

Without loss of generality, for group $z$, we assume the proposed $\mc S$-score variable $\mc S_z$ takes values in a set of discrete values $\{1,2 \dots, M\}$, with corresponding probability masses $p^z_1, p^z_2, \dots, p^z_M$. {Here $M=h+\tau+1$ for notation convenience in the following derivation. } This means we derive the variance based on assuming both $Y_1$ and $Y_2$ variables are discrete, although both of them can be continuous. This is because assuming a finite support allows us to parameterize the problem using a finite number of parameters, and thus we can use the M-estimation theory to construct the estimator and derive the asymptotic variance. In practice, when the observed survival time or non-survival variables are continuous, our derivation and results can still apply by treating the finite observations in the sample as the total number of parameters. 

The survival function of $\mc S$-score for group $z$ is denoted by $\bar{F}^z_s = P(\mc S_z > s).$ We let $p_0=0$ and $\bar{F}^z_M=0$, where $z\in\{a,b\}$. 
With these new definitions, the above \eqref{eq:ND} can be re-written as 
\begin{align*}
    N = \sum_{s=1}^M(\bar F_{s-1}^b-\bar F_s^b)\bar F_s^a,\quad  D = \sum_{s=1}^M(\bar F_{s-1}^a-\bar F_s^a)\bar F_s^b. 
\end{align*}
Moreover, we consider the parametrization based on conditional survival probabilities, where we let 
\begin{align*}
    \bar F_s^z = \prod_{j=1}^s q_j^z, \quad s=1,\dots,M-1.
\end{align*}
with $\bar F_0^z=1$ and $\bar F_M^z=0$, for both $z\in\{a,b\}$. Here, $q_j^z=P(\mc S_z>j\mid \mc S_z\geq j-1) = \bar F_j^z/\bar F_{j-1}^z$ is the conditional survival probability at time $j$ conditioning that the participant survived beyond $j-1$, for both $z\in\{a,b\}$. 

Then, the conditional survival probabilities $\bd q = (q_1^a,\dots,q_{M-1}^a,q_1^b,\dots,q_{M-1}^b)$ can be estimated via the following estimating equations:
\begin{align*}
    \frac1n\sum_{i=1}^n \psi_s^z(\mc O^{\mc S}_i) = 0, 
\end{align*}
where for a copy of the observed data $\mc O^{\mc S}$, we denote
\begin{align*}
    \psi_s^z = \psi_s^z(\mc O^{\mc S}) = I(Z=z)E_{s}(A_{s}-\widehat q_s^z),
\end{align*}
where $s\in\{1,\dots,M-1\}$ and $z\in\{a,b\}$, where the solution $\widehat q_s^z$ corresponds to the Kaplan-Meier estimator for $q_s^z$ at $\mc S_z=s$. 
$E_s$ is the indicator of whether the individual is in the risk set at $\mc S=s$ and $A_s$ is the indicator of whether the individual survives beyond time $\mc S=s$.  We stack these estimating equations into the vector
$$
\Psi(\bd q) = \left(  \psi^a_1, \dots,  \psi^a_{M-1}, \psi^b_1, \dots, \psi^b_{M-1} \right)'.
$$
Let $B(\bd q) = -\dfrac{\partial\Psi(\bd q)}{\partial\bd q}$, which is diagonal matrix since each equation in the above vector depends only on one of elements in $\bd q$, i.e., $q_s^z$. For group $z$ the derivative with respect to $q_s^z$ is 
\begin{align*}
    \frac{\partial\psi_s^z}{\partial q_s^z} = -I(Z=z)E_s.
\end{align*}
Therefore, the (population) diagonal entry of $B(\bd q)$ is $\Ex\{I(Z=z)E_s\}$. Therefore, following \cite{van2000asymptotic}, the influence function (IF) for these $q$'s is 
\begin{align*}
   \eta = \eta(\mc O^{\mc S}) = B(\bd q)^{-1}\psi,
\end{align*}
where $\mc O^{\mc S}$ is the observed data defined in Section \ref{subsec:propose-obs}. Empirically, one estimates these by replacing the expectation to the sample mean. That is 
\begin{align*}
    \widehat\eta(\mc O^{\mc S}) = \frac{I(Z=z)E_s(A_s-\widehat q_s^z)}{\widehat\Ex\{I(Z=z)E_s\}}. 
\end{align*}

Next, to get the IF of the NPMLE \eqref{eq:WR-npmle} for $\WR$, we need to further obtain the following derivative:
$$
C = \left( \frac{\partial \WR}{\partial q^a_1}, \dots, \frac{\partial \WR}{\partial q^a_{M-1}}, \frac{\partial \WR}{\partial q^b_1}, \dots, \frac{\partial \WR}{\partial q^b_{M-1}} \right).
$$
For each element $\dfrac{\partial\WR}{\partial q^z_t}$,  we have 
$\displaystyle
\frac{\partial\WR}{\partial q^z_s} = \sum_{k=s}^{M-1}\frac{\partial\WR}{\partial\bar F^z_k}\frac{\partial\bar F^z_k}{\partial q^z_s}  ~~\text{ (by the chain rule)}. 
$

Note that all $\bar F^z_k$, where $k\geq s$,  dependent on $q_s^z$ for a given $z\in\{a,b\}$. 
We calculate  $\dfrac{\partial\bar F^z_k}{\partial q^z_s}$ as follows. Since,
$
\displaystyle\log \bar F^z_k =\sum_{j=1}^{k}\log q^z_j,
$
differentiating with respect to $q^z_s$ for $s\leq k$, gives us
$\displaystyle
\frac{\partial\log\bar F^z_k}{\partial q^z_s}=\frac{1}{q^z_s}.
$ Thus, by the chain rule,
$$
\frac{\partial\bar F^z_k}{\partial q^z_s}=\bar F^z_k\frac{1}{q^z_s} = \frac{\bar F^z_k}{q^z_s}, \quad t\leq k.
$$ 
For $s>k$, the derivatives are zero since $\bar F^a_k$ and $\bar F^b_k$ do not depend on $q^a_s$ or $q^b_s$ beyond time $k$.

Additionally, for $\dfrac{\partial\WR}{\partial\bar F^z_s}$, applying the quotient rule,
$$
\frac{\partial \WR}{\partial\bar F^a_s} = \frac{D (\bar F^b_{s-1} - \bar F^b_s) - N (\bar F^b_s - \bar F^b_{s+1})}{D^2}
~~
\text{ and }~~
\frac{\partial \WR}{\partial\bar F^b_s} = \frac{D (\bar F^a_s - \bar F^a_{s+1}) - N (\bar F^a_{s-1} - \bar F^a_s)}{D^2},
$$
where $N$ and $D$ are the numerator and denominator defined in \eqref{eq:ND}. 

Finally, by Delta method, the IF for the NPMLE \eqref{eq:WR-npmle} is given as
\begin{align*}
\varphi(\mc O^{\mc S}) & = C\eta(\mc O^{\mc S}) \\
& = \sum_{t=1}^{M-1} \frac{1}{q^a_s} \sum_{k=s}^{M-1} \bar F^a_k D^{-2} 
\bigg[ D (\bar F^b_{k-1} - \bar F^b_k) - N (\bar F^b_k - \bar F^b_{k+1}) \bigg] 
\frac{I(Z=a)E_s(A_s - q^a_s)}{\Ex\{I(Z=a)E_s\}} \\
& \qquad
+ \sum_{t=1}^{M-1} \frac{1}{q^b_s} \sum_{k=s}^{M-1} \bar F^b_k D^{-2} 
\bigg[ D (\bar F^a_k - \bar F^a_{k+1}) - N (\bar F^a_{k-1} - \bar F^a_k) \bigg] 
\frac{I(Z=b)E_s(A_s - q^b_s)}{\Ex\{I(Z=b)E_s\}}.
\end{align*}
To estimate $\varphi(\mc O^{\mc S})$, one can plug-in the estimated survival and conditional survival functions and replace the expectation by sample mean, resulting the estimated IF $\widehat\varphi(\mc O^{\mc S})$. 
Therefore, the estimated variance based on the IF can be given by 
\begin{align*}
    \widehat{\mc V}_{\text{IF}} = \frac1n\sum_{i=1}^n \widehat\varphi(\mc O^{\mc S}_i)^2.
\end{align*}
In the finite-support setting considered here, the NPMLE is regular and asymptotically linear (RAL) by standard M-estimation theory.

The estimator $\WRhat$ is consistent and $n^{-1}\widehat{\mc V}_{\text{IF}}$ is its estimated asymptotic variance, i.e., 
\begin{align*}
    \sqrt{n/\widehat{\mc V}_{\text{IF}}}(\WRhat-\WR)\xrightarrow{d}\mc N(0,1). 
\end{align*}

\subsection{Derivation and properties of the covariate-adjusted estimator}\label{subapp:coad}

Consider that both $Y_1$ and $Y_2$ are discrete variables, and without loss of generality, we assume the support of $\mc S$-score is $\{0,1,\dots,h,h+1,\dots,h+\tau+1\}$ for some positive $h<\infty$ (time horizon) and $\tau=\max(Y_2)<\infty$. $F_z(s)$, $F_{1z}(s)$ and $F_{2z}(s)$ denote, respectively, the CDFs of $\mc S$, $Y_1$ and $Y_2$ at  $s$ under the treatment group $Z=z$, for $z\in\{a,b\}$. 

For any $s \leq h$, we have $F_z(s) = F_{1z}(s)$, so the Kaplan-Meier estimator can be directly applied. For $s > h$, note that $\mc S = h + 1 + Y_2 = h + 1 + (s - h - 1) = s$, which corresponds to the event that $Y_1 > h$ and $Y_2 = s - h - 1$. Therefore, the CDF of $\mc S$ at $s$ is given by $F_z(s) = F_{1z}(h) + {1 - F_{1z}(h)} F_{2z}(s - h - 1)$. Accordingly, our goal is to estimate $F_{1z}(s)$ for $s \in \{1, \dots, h\}$ using the Kaplan-Meier estimator, and $F_{2z}(s)$ for $s \in \{1, \dots, \tau\}$ using a covariate-adjusted estimator. 

To this end, we focus on estimating the parameter vector $(F_{1z}(1), \dots, F_{1z}(h), \bd\beta_z, F_{2z}(1), \dots, F_{2z}(\tau))$, where $\bd\beta_z$ parameterizes the missingness model for $R_{2z}$. Specifically, we define $\pi_z(\mb X) = P(R_{2z} = 1 \mid \mb X)$ as the propensity score for observing $Y_2$, conditional on covariates $\mb X$. Under Assumption \ref{asp:MAR-X}, $\pi_z(\mb X)$ serves as a summary score for a participant's probability of having observed $Y_2$, given their baseline covariates. Since $\pi_z(\mb X)$ is generally unknown, we specify a parametric regression model, commonly logistic, of the form $\pi_z(\mb X; \bd\beta_z) = \{1 + \exp(-\mb X'\bd\beta_z)\}^{-1}$, where $\bd\beta_z$ is typically of the same dimension as $\mb X$. The set of unbiased estimating equations for the above parameter vector is specified by 
\begin{align}\label{eq:ipw-esteq}
    \frac1n\sum_{i=1}^n
    I(Z_i=z)\begin{pmatrix}
    \Delta_{1i}\left\{I(Y_{1i}\leq 1)-F_{1z}(1)\right\}\\
        \vdots \\
    \Delta_{1i}\left\{I(Y_{1i}\leq h)-F_{1z}(h)\right\}\\
    \phi_z(\bd\beta_z; R_{2i}, \mb X_i)\\
    R_{2i}\pi_z(\mb X_i;\bd\beta_z)^{-1}\left\{I(Y_{2i}\leq 1)-F_{2z}(1)\right\}\\
        \vdots \\
    R_{2i}\pi_z(\mb X_i;\bd\beta_z)^{-1}\left\{I(Y_{2i}\leq \tau)-F_{2z}(\tau)\right\}
    \end{pmatrix} = 0, 
\end{align}
where the first $h$ terms defined the Kaplan-Meier estimator for the CDF of $Y_{1z}$, 
$\phi_z(\bd\beta_z; R_2,\mb X)$ is the estimating equation for $\bd\beta_z$ (e.g., the score equation for the logistic regression model), and the last $\tau$ terms defined the IPW-weighted estimator for CDF of $Y_2$.  

Let $(\widehat F_{1z}(1), \dots, \widehat F_{1z}(h), \widehat{\bd\beta}_z, \widehat F_{2z}(1), \dots, \widehat F_{2z}(\tau))$ denote the solution to the above system of estimating equations. Then, the resulting estimator for the CDF of the $\mc S$-score is then given by 
\begin{align}\label{eq:coad-est-S}
    \widehat F_z(s) = I(s\leq h)\cdot\widehat F_{1z}(s) + I(s>h)\cdot[\widehat F_{1z}(h) + \{1-\widehat F_{1z}(h)\}\widehat F_{2z}(s-h-1)],
\end{align}
for $s\in\{0,1,\dots,h,h+1,\dots,h+\tau+1\}$ with $\widehat F_z(0)=0$. Plugging this covariate-adjusted estimator into \eqref{eq:WR-npmle-obs} yields the covariate-adjusted estimator for the WR, referred to as the semiparametric estimator in Section \ref{subsec:covar}. 

Thus far, we have derived the covariate-adjusted estimator under the setting where both the survival time and the non-survival endpoint take discrete values. If at least one of the endpoints is continuous, the corresponding M-estimating equations in \eqref{eq:ipw-esteq} become infinite-dimensional, introducing both theoretical and computational challenges. {A practical workaround is to coarsen the continuous outcomes into discrete categories---for example, by rounding exact times to the nearest day or month.} However, extending the covariate-adjusted estimator and its asymptotic variance estimator to fully continuous settings remains an open question for future research. 

\subsubsection{Consistency of the covariate-adjusted estimator}

To establish the consistency, we follow a similar argument as in Appendix \ref{subapp:proof-consistency}. Once we establish the convergence of the covariate-adjusted estimators of the CDFs (i.e., $\widehat F_{1z}(s)\xrightarrow{p} F_{1z}(s)$ for s$\in[0,h]$ and $\widehat F_{2z}(s)\xrightarrow{p} F_{2z}(s)$ for $s\in[0,\tau]$, for both $z\in\{a,b\}$), we can show the consistency of the covariate-adjusted WR estimator to the true WR. 

We assume the propensity score model is correctly specified (due to Assumption \ref{asp:MAR-X}) in the sense that the estimated parameter converges to a probability limit $\widehat{\bd\beta}_z\xrightarrow{p}\bd\beta_z^*$, and $\pi_z(\mb X) = \pi_z(\mb X;\bd\beta_z^*)$ is the true propensity score, for $z\in\{a,b\}$. 

The uniform convergence of $\widehat F_{1z}(s)$ over $s\in[0,h]$ is proved in Section \ref{subapp:proof-consistency}; we only need to focus on $\widehat F_{2z}(s)$ for any $s\in[0,\tau]$. Note that using the estimating equations \eqref{eq:ipw-esteq} we have
$$
\widehat F_{2z}(s) = \frac{n^{-1}\sum_{i=1}^n I(Z_i=z)R_{2i}\pi_z(\mb X_i;\widehat{\bd\beta}_z)^{-1}I(Y_{2i}\leq s)}{n^{-1}\sum_{i=1}^n I(Z_i=z)R_{2i}\pi_z(\mb X_i;\widehat{\bd\beta}_z)^{-1}} : = \frac{I_1}{I_2}.
$$
The numerator $I_1  = \displaystyle\frac1n \displaystyle\sum_{i=1}^n I(Z_i=z)\frac{R_{2i}}{\pi_z(\mb X_i;\widehat{\bd\beta}_z)}I(Y_{2i}\leq s)  \xrightarrow{p}\Ex\left\{I(Z=z)\frac{R_2}{\pi_z(\mb X)}I(Y_2\leq s)\right\},$ where

\allowdisplaybreaks\begin{align*}
    \Ex\left\{I(Z=z)\frac{R_2}{\pi_z(\mb X)}I(Y_2\leq s)\right\} & = \Ex\left[\Ex\left\{I(Z=z)\frac{R_2}{\pi_z(\mb X)}I(Y_2\leq s)~\bigg|~\mb X, Z=z\right\}\right] \\
    & = \Ex\left[\frac{I(Z=z)}{\pi_z(\mb X)}\Ex\left\{R_2I(Y_2\leq s)\mid\mb X,Z=z\right\}\right] \\
    & = \Ex\left[\frac{I(Z=z)}{\pi_z(\mb X)}\Ex\{I(Y_2\leq s)\mid\mb X,Z=z\}\Px(R_2=1\mid\mb X,Z=z)\right] ~~\text{(by Assumption \ref{asp:MAR-X})}
    \\
    & = \Ex\left[\frac{I(Z=z)}{\pi_z(\mb X)}\pi_z(\mb X)\Ex\{I(Y_2\leq s)\mid\mb X,Z=z\}\right]  = \Px(Z=z, Y_2\leq s). 
\end{align*}
 Also, $I_2  =  \displaystyle\frac1n \displaystyle\sum_{i=1}^n I(Z_i=z)\frac{R_{2i}}{\pi_z(\mb X_i;\widehat{\bd\beta}_z)} \xrightarrow{p}\Ex\left\{I(Z=z)\frac{R_2}{\pi_z(\mb X)}\right\}$ with
\begin{align*}
   \Ex\left\{I(Z=z)\frac{R_2}{\pi_z(\mb X)}\right\} & = \Ex\left[\Ex\left\{I(Z=z)\frac{R_2}{\pi_z(\mb X)}~\bigg|~\mb X, Z=z\right\}\right] = \Ex\left[\frac{I(Z=z)}{\pi_z(\mb X)}\Ex\left\{R_2\mid\mb X,Z=z\right\}\right] \\
    & = \Ex\left[\frac{I(Z=z)}{\pi_z(\mb X)}\Px(R_2=1\mid\mb X,Z=z)\right] = \Ex\left[\frac{I(Z=z)}{\pi_z(\mb X)}\pi_z(\mb X)\right] = \Px(Z=z). 
\end{align*}
Thus, 
\begin{align*}
    \widehat F_{2z}(s)\xrightarrow{p}\frac{\Px(Z=z, Y_2\leq s)}{\Px(Z=z)} = \Px(Y_2\leq s\mid Z=z) = F_{2z}(s). 
\end{align*}
Since the proof is the same for any $s\in[0,\tau]$, with $0<\tau<\infty$, we can construct the uniform consistency of $\widehat F_{2z}(s)$ over $s$. Then, the uniform consistency of the estimator \eqref{eq:coad-est-S} over $\mc S\in[0,h+\tau+1]$ follows immediately.

\subsubsection{Variance estimation of the covariate-adjusted estimator}

In addition to the property of consistency, we derive the asymptotic variance of the estimator \eqref{eq:coad-est-S}, following a similar derivation as in Appendix \ref{subapp:varest}. We also consider that the $\mc S$-score $\mc S_z$ for the group $z$ takes values in the set $\{1,\dots,h,h+1,\dots,h+\tau+1\}$, where $h$ and $\tau$ are some positive integers. For $\mc S_z\leq h$, we have $\mc S_z=Y_{1z}$. After $h$, $\mc S_z=1+h+Y_{2z}$, and we have IPW weights on values after $h$. 

Since the propensity score is parametrized by $\bd\beta_z$, we consider the vector of parameters as conditional survival probabilities and $(\bd\beta_a',\bd\beta_b')'$ as $\bd\theta=(\bd\beta_a', q_1^a,\dots,q_{M-1}^a,\bd\beta_b',q_1^b,\dots,q_{M-1}^b)'$, where $M=h+\tau+1$. The parameter-vector $\bd\theta$ can jointly be estimated by solving the following equations
\begin{align*}
    \frac1n\sum_{i=1}^n\phi_z(\mc O^{\mc S}) = 0 \quad \text{and}\quad \frac1n\sum_{i=1}^n\psi_s^z(\mc O^{\mc S}_i) = 0,
\end{align*}
where for a copy of the observed data $\mc O^{\mc S}$, we have
\allowdisplaybreaks\begin{align*}
    \phi_z=\phi_z(\mc O^{\mc S})  & = I(Z=z)\phi_z(\widehat{\bd\beta}_z; R_{2}, \mb X) = 0,\\
    \psi_s^z=\psi_s^z(\mc O^{\mc S}) & = I(Z=z)E_s(A_s-\widehat q_s^z) = 0 ~~\text{ for }s\leq h, ~~\text{ and} \\
    \psi_s^z=\psi_s^z(\mc O^{\mc S}) & = I(Z=z)\pi_z(\mb X; \widehat{\bd\beta}_z)^{-1}E_s(A_s-\widehat q_s^z) = 0 ~~\text{ for }s>h, 
\end{align*}
As an illustration, we consider the special case where we use logistic regression to estimate the propensity score (which is a common choice in practice). At the end, we will show that the choice of the propensity score model does not affect the expression of the variance formula, because the propensity score does not play a role in the definition of the $\WR$ estimand, and it is only a nuisance parameter. 

Let $\pi_z(\mb X;\bd\beta_z)=\{1+\exp(-\mb X'\bd\beta_z)\}^{-1}$, the propensity score estimated via logistic regression and define
   $ \phi_z(\bd\beta_z; R_2, \mb X)  = \{R_2-\pi_z(\mb X;\bd\beta_z)\}\mb X, $ for $z\in\{a,b\}$.
We stack these estimating equations into the vector 
$$
\Psi(\bd q) = \left(\phi_a,  \psi^a_1, \dots,  \psi^a_{M-1}, \phi_b, \psi^b_1, \dots, \psi^b_{M-1} \right)'.
$$
Let $B(\bd\theta) = -\dfrac{\partial\Psi(\bd\theta)}{\partial\bd\theta}$, which is diagonal in multiple blocks since each equation in the above vector depends only on one of the unique elements in $\bd\theta$. For the group $z$, the derivative is computed by
\begin{align*}
    \frac{\partial\phi_z}{\partial\bd\beta_z} & = -\pi_z(\mb X;\bd\beta_z)\{1-\pi_z(\mb X;\bd\beta_z)\}\mb X\mb X' \quad\text{(specific to logistic regression)}, \\
    \frac{\partial\psi_s^z}{\partial q_s^z} & = -I(Z=z)E_s \quad\text{for }s\leq h, \quad\text{and} \\
    \frac{\partial\psi_s^z}{\partial q_s^z} & = -I(Z=z)\pi_z(\mb X;\bd\beta_z)^{-1}E_s \quad \text{for }t>h. 
\end{align*} 
Following \cite{van2000asymptotic}, the IF for $\bd\theta$ is
\begin{align*}
   \eta = \eta(\mc O^{\mc S}) = B(\bd\theta)^{-1}\Psi(\bd\theta),
\end{align*}
where $\mc O^{\mc S}$ is the observed data with covariates $\mb X$. 

Next, to get the IF of the covariate-adjusted estimator \eqref{eq:coad-est-S} for $\WR$, we need to further obtain the following derivative:
$$
C = \left( \frac{\partial\WR}{\partial\bd\beta_a'}, \frac{\partial \WR}{\partial q^a_1}, \dots, \frac{\partial \WR}{\partial q^a_{M-1}}, \frac{\partial\WR}{\partial\bd\beta_b'}, \frac{\partial \WR}{\partial q^b_1}, \dots, \frac{\partial \WR}{\partial q^b_{M-1}}\right)'.
$$
Notice that
$$
\frac{\partial\WR}{\partial\bd\beta_z} = \mb 0,
$$
as the definition of the WR does not involve the propensity score. The propensity score is only an intermediate nuisance function we derived in the estimation process. 

Other derivations follow exactly the same as those in Section \ref{subapp:varest}. Finally, the IF of the covariate-adjusted estimator \eqref{eq:coad-est-S} is given as
\begin{align*}
\varphi(\mc O^{\mc S}) & = C\eta(\mc O^{\mc S}) \\
& = \sum_{s=1}^{M-1} \frac{1}{q^a_s} \sum_{k=s}^{M-1} \bar F^a_k D^{-2} 
\bigg[ D (\bar F^b_{k-1} - \bar F^b_k) - N (\bar F^b_k -\bar F^b_{k+1}) \bigg] 
\frac{I(Z=a)\{I(s\leq h)+I(s>h)\pi_a(\mb X;\bd\beta_a)^{-1}\}E_s (A_s - q^a_s)}{\Ex[I(Z=a)\{I(s\leq h)+I(s>h)\pi_a(\mb X;\bd\beta_a)^{-1}\}E_s]} \\
& \quad
+ \sum_{s=1}^{M-1} \frac{1}{q^b_s} \sum_{k=s}^{M-1}\bar F^b_k D^{-2} 
\bigg[ D (\bar F^a_k - \bar F^a_{k+1}) - N (\bar F^a_{k-1} - \bar F^a_k) \bigg] 
\frac{I(Z=b)\{I(s\leq h)+I(s>h)\pi_b(\mb X;\bd\beta_b)^{-1}\}E_s (A_s - q^a_s)}{\Ex[I(Z=b)\{I(s\leq h)+I(s>h)\pi_b(\mb X;\bd\beta_b)^{-1}\}E_s]},
\end{align*}
where notation $N$, $D$, $\bar F_k^a$, $\bar F_k^b$ are defined in Section \ref{subapp:varest}. Then, we can consistently estimate the asymptotic variance by the mean square of the estimated IFs. 

\newpage

\begin{center}
    \LARGE Online Supplemental Material
\end{center}

\appendix\label{Appendix}

\setcounter{section}{0}
\renewcommand{\thesection}{S.\arabic{section}}
\counterwithin{equation}{section}
\counterwithin{table}{section}
\counterwithin{figure}{section}

\section{Complete Simulation Results}\label{app:simu}

\subsection{Tables for full simulation settings}

\begin{table}[H]
\begin{threeparttable}
    \centering
    \scriptsize
    \begin{tabular}{cclrrrrrrrrrrrrrrrrrrr}
    \toprule
    & & & \multicolumn{4}{c}{$\WR=1$} & \multicolumn{4}{c}{$\WR=2$}\\
    \makecell[l]{$(n_a, n_b)$} & \makecell[c]{Missing \\ data of $Y_2$} & Method  & ARB\% & RMSE & CP\% & Width & ARB\% & RMSE & CP\% & Width \\
    \cmidrule(lr){1-3}\cmidrule(lr){4-7}\cmidrule(lr){8-11}
   &  & $\mc S$-score (IF-Wald) & 1.78 & 0.166 & 95.55 & 0.66 & 2.69 & 0.390 & 94.85 & 1.48 \\ 
   & No & $\mc S$-score (BT-Wald) & 1.78 & 0.166 & 96.25 & 0.68 & 2.69 & 0.390 & 95.60 & 1.55 \\ 
   &  & $\mc S$-score (BT-QT) & 1.78 & 0.166 & 95.10 & 0.67 & 2.69 & 0.390 & \textbf{93.55} & 1.54 \\ 
   &  & Pocock & 1.78 & 0.166 & 95.40 & 0.67 & 2.69 & 0.390 & 95.10 & 1.51 \\ 
   \addlinespace
   &  & $\mc S$-score (IF-Wald) & 1.88 & 0.168 & 94.95 & 0.66 & 1.68 & 0.369 & 95.00 & 1.46 \\ 
   & MCAR & $\mc S$-score (BT-Wald) & 1.88 & 0.168 & 95.50 & 0.68 & 1.68 & 0.369 & 95.80 & 1.54 \\ 
   & (20\%) & $\mc S$-score (BT-QT) & 1.88 & 0.168 & 94.65 & 0.67 & 1.68 & 0.369 & 94.70 & 1.52 \\ 
   &  & Pocock & 1.93 & 0.171 & 95.35 & 0.68 & 3.15 & 0.358 & 94.25 & 1.43 \\ 
   \addlinespace
   &  & $\mc S$-score (IF-Wald) & 1.65 & 0.170 & 94.75 & 0.66 & 2.06 & 0.391 & 94.25 & 1.46 \\ 
  (100, 100) & MCAR & $\mc S$-score (BT-Wald) & 1.65 & 0.170 & 95.40 & 0.68 & 2.06 & 0.391 & 95.30 & 1.54 \\ 
  Small & (40\%) & $\mc S$-score (BT-QT) & 1.65 & 0.170 & 94.45 & 0.67 & 2.06 & 0.391 & \textbf{93.95} & 1.52 \\ 
   &  & Pocock & 1.76 & 0.175 & 95.25 & 0.70 & 7.72 & 0.392 & \textbf{90.25} & 1.37 \\
   \addlinespace
   &  & $\mc S$-score (IF-Wald) & 1.78 & 0.171 & 95.20 & 0.66 & 2.13 & 0.392 & 94.90 & 1.47 \\ 
   & MAR & $\mc S$-score (BT-Wald) & 1.78 & 0.171 & 95.40 & 0.68 & 2.13 & 0.392 & 95.80 & 1.54 \\ 
   & (20\%) & $\mc S$-score (BT-QT) & 1.78 & 0.171 & 94.00 & 0.67 & 2.13 & 0.392 & 94.95 & 1.53 \\ 
   &  & Pocock & 5.82 & 0.169 & \textbf{93.30} & 0.63 & 6.39 & 0.382 & \textbf{92.60} & 1.38 \\
   \addlinespace
   &  & $\mc S$-score (IF-Wald) & 2.20 & 0.173 & 94.85 & 0.66 & 2.19 & 0.373 & 95.45 & 1.47 \\ 
   & MAR & $\mc S$-score (BT-Wald) & 2.20 & 0.173 & 95.30 & 0.68 & 2.19 & 0.373 & 96.40 & 1.54 \\ 
   & (40\%) & $\mc S$-score (BT-QT) & 2.20 & 0.173 & \textbf{93.60} & 0.67 & 2.19 & 0.373 & 94.95 & 1.53 \\ 
   &  & Pocock & 13.12 & 0.200 & \textbf{85.80} & 0.60 & 14.93 & 0.441 & \textbf{84.05} & 1.27 \\ 
   \addlinespace
   \midrule
   &  & $\mc S$-score (IF-Wald) & 0.37 & 0.052 & 95.00 & 0.20 & 0.30 & 0.114 & 95.00 & 0.45 \\ 
   & No & $\mc S$-score (BT-Wald) & 0.37 & 0.052 & 94.50 & 0.20 & 0.30 & 0.114 & 95.05 & 0.45 \\ 
   &  & $\mc S$-score (BT-QT) & 0.37 & 0.052 & 94.00 & 0.20 & 0.30 & 0.114 & 94.35 & 0.45 \\ 
   &  & Pocock & 0.37 & 0.052 & 94.65 & 0.20 & 0.30 & 0.114 & 94.90 & 0.45 \\
   \addlinespace
   &  & $\mc S$-score (IF-Wald) & 0.04 & 0.053 & 94.70 & 0.20 & 0.06 & 0.113 & 95.10 & 0.45 \\ 
   & MCAR & $\mc S$-score (BT-Wald) & 0.04 & 0.053 & 94.65 & 0.20 & 0.06 & 0.113 & 95.05 & 0.45 \\ 
   & (20\%) & $\mc S$-score (BT-QT) & 0.04 & 0.053 & 94.25 & 0.20 & 0.06 & 0.113 & 94.95 & 0.45 \\ 
   &  & Pocock & 0.03 & 0.054 & 94.40 & 0.21 & 4.64 & 0.145 & \textbf{86.15} & 0.43 \\ 
   \addlinespace
   &  & $\mc S$-score (IF-Wald) & 0.10 & 0.052 & 95.05 & 0.20 & 0.01 & 0.113 & 95.45 & 0.45 \\ 
  (1000, 1000) & MCAR & $\mc S$-score (BT-Wald) & 0.10 & 0.052 & 94.80 & 0.20 & 0.01 & 0.113 & 95.50 & 0.45 \\ 
  Large & (40\%) & $\mc S$-score (BT-QT) & 0.10 & 0.052 & 94.50 & 0.20 & 0.01 & 0.113 & 95.00 & 0.45 \\ 
   &  & Pocock & 0.11 & 0.054 & 95.45 & 0.21 & 9.67 & 0.226 & \textbf{53.30} & 0.41 \\
   \addlinespace
   &  & $\mc S$-score (IF-Wald) & 0.51 & 0.054 & 94.05 & 0.20 & 0.05 & 0.113 & 95.25 & 0.45 \\ 
   & MAR & $\mc S$-score (BT-Wald) & 0.51 & 0.054 & 94.20 & 0.20 & 0.05 & 0.113 & 95.25 & 0.45 \\ 
   & (20\%) & $\mc S$-score (BT-QT) & 0.51 & 0.054 & 93.70 & 0.20 & 0.05 & 0.113 & 95.00 & 0.45 \\ 
   &  & Pocock & 6.69 & 0.084 & \textbf{73.10} & 0.19 & 8.26 & 0.201 & \textbf{64.20} & 0.42 \\ 
   \addlinespace
   &  & $\mc S$-score (IF-Wald) & 0.48 & 0.051 & \textbf{96.40} & 0.20 & 0.08 & 0.110 & 95.90 & 0.45 \\ 
   & MAR & $\mc S$-score (BT-Wald) & 0.48 & 0.051 & \textbf{96.40} & 0.20 & 0.08 & 0.110 & 95.85 & 0.45 \\ 
   & (40\%) & $\mc S$-score (BT-QT) & 0.48 & 0.051 & 95.85 & 0.20 & 0.08 & 0.110 & 95.70 & 0.45 \\ 
   &  & Pocock & 14.57 & 0.152 & \textbf{15.95} & 0.18 & 16.80 & 0.362 & \textbf{8.70} & 0.38 \\
   \addlinespace
  \bottomrule
    \end{tabular}
    \caption{Simulation results: no censoring on $Y_1$.}
    \label{tab:simu-nocen}
    \begin{tablenotes}
	\scriptsize
	\item CP\%'s outside the interval [94, 96] are highlighted in bold.
     \end{tablenotes}
\end{threeparttable}
\end{table}

\begin{table}[H]
\begin{threeparttable}
    \centering
    \scriptsize
    \begin{tabular}{cclrrrrrrrrrrrrrrrrrrr}
    \toprule
    & & & \multicolumn{4}{c}{$\WR=1$} & \multicolumn{4}{c}{$\WR=2$}\\
    \makecell[l]{$(n_a, n_b)$} & \makecell[c]{Missing \\ data of $Y_2$} & Method  & ARB\% & RMSE & CP\% & Width & ARB\% & RMSE & CP\% & Width \\
    \cmidrule(lr){1-3}\cmidrule(lr){4-7}\cmidrule(lr){8-11}
    &  & $\mc S$-score (IF-Wald) & 1.70 & 0.180 & 95.20 & 0.68 & 2.34 & 0.396 & 94.90 & 1.51 \\ 
   & No & $\mc S$-score (BT-Wald) & 1.70 & 0.180 & 95.85 & 0.71 & 2.34 & 0.396 & 95.80 & 1.60 \\ 
   &  & $\mc S$-score (BT-QT) & 1.70 & 0.180 & 94.65 & 0.70 & 2.34 & 0.396 & 94.40 & 1.58 \\ 
   &  & Pocock & 1.89 & 0.189 & 95.00 & 0.73 & 2.24 & 0.411 & 94.95 & 1.60 \\ 
   \addlinespace
   &  & $\mc S$-score (IF-Wald) & 1.52 & 0.177 & 94.35 & 0.68 & 1.54 & 0.387 & 94.95 & 1.50 \\ 
   & MCAR & $\mc S$-score (BT-Wald) & 1.52 & 0.177 & 94.90 & 0.71 & 1.54 & 0.387 & 95.45 & 1.58 \\ 
   & (20\%) & $\mc S$-score (BT-QT) & 1.52 & 0.177 & 94.35 & 0.70 & 1.54 & 0.387 & 94.60 & 1.56 \\ 
   &  & Pocock & 1.74 & 0.189 & 95.05 & 0.74 & 2.49 & 0.394 & \textbf{93.70} & 1.53 \\ 
   \addlinespace
   &  & $\mc S$-score (IF-Wald) & 1.64 & 0.178 & 94.85 & 0.69 & 1.90 & 0.381 & 95.75 & 1.50 \\ 
  (100, 100) & MCAR & $\mc S$-score (BT-Wald) & 1.64 & 0.178 & 95.40 & 0.71 & 1.90 & 0.381 & 95.95 & 1.58 \\ 
  Small & (40\%) & $\mc S$-score (BT-QT) & 1.64 & 0.178 & 95.15 & 0.70 & 1.90 & 0.381 & 95.40 & 1.56 \\ 
   &  & Pocock & 2.36 & 0.195 & 95.35 & 0.76 & 6.11 & 0.390 & \textbf{92.95} & 1.49 \\ 
   \addlinespace
   &  & $\mc S$-score (IF-Wald) & 1.49 & 0.175 & 94.65 & 0.68 & 1.97 & 0.398 & 94.65 & 1.50 \\ 
   & MAR & $\mc S$-score (BT-Wald) & 1.49 & 0.175 & 95.35 & 0.71 & 1.97 & 0.398 & 95.35 & 1.59 \\ 
   & (20\%) & $\mc S$-score (BT-QT) & 1.49 & 0.175 & 95.25 & 0.70 & 1.97 & 0.398 & 94.40 & 1.57 \\ 
   &  & Pocock & 4.69 & 0.181 & \textbf{93.90} & 0.70 & 4.93 & 0.402 & \textbf{93.30} & 1.50 \\ 
   \addlinespace
   &  & $\mc S$-score (IF-Wald) & 2.38 & 0.176 & 95.70 & 0.69 & 1.43 & 0.390 & 94.65 & 1.49 \\ 
   & MAR & $\mc S$-score (BT-Wald) & 2.38 & 0.176 & \textbf{96.20} & 0.72 & 1.43 & 0.390 & 95.30 & 1.58 \\ 
   & (40\%) & $\mc S$-score (BT-QT) & 2.38 & 0.176 & 95.25 & 0.71 & 1.43 & 0.390 & 94.80 & 1.56 \\ 
   &  & Pocock & 10.95 & 0.199 & \textbf{89.20} & 0.66 & 12.44 & 0.435 & \textbf{87.30} & 1.39 \\
   \addlinespace
   \midrule
   &  & $\mc S$-score (IF-Wald) & 0.18 & 0.054 & 94.15 & 0.21 & 0.16 & 0.113 & 95.60 & 0.46 \\ 
   & No & $\mc S$-score (BT-Wald) & 0.18 & 0.054 & 94.45 & 0.21 & 0.16 & 0.113 & 95.90 & 0.46 \\ 
   &  & $\mc S$-score (BT-QT) & 0.18 & 0.054 & \textbf{93.80} & 0.21 & 0.16 & 0.113 & 95.40 & 0.46 \\ 
   &  & Pocock & 0.18 & 0.056 & 94.90 & 0.22 & 0.15 & 0.117 & 95.95 & 0.48 \\ 
   \addlinespace
   &  & $\mc S$-score (IF-Wald) & 0.09 & 0.053 & 95.15 & 0.21 & 0.25 & 0.119 & 94.75 & 0.46 \\ 
   & MCAR & $\mc S$-score (BT-Wald) & 0.09 & 0.053 & 95.15 & 0.21 & 0.25 & 0.119 & 94.95 & 0.46 \\ 
   & (20\%) & $\mc S$-score (BT-QT) & 0.09 & 0.053 & 94.55 & 0.21 & 0.25 & 0.119 & 94.45 & 0.46 \\ 
   &  & Pocock & 0.13 & 0.058 & 94.80 & 0.23 & 3.84 & 0.143 & \textbf{88.75} & 0.46 \\
   \addlinespace
   &  & $\mc S$-score (IF-Wald) & 0.18 & 0.054 & 95.15 & 0.21 & 0.06 & 0.119 & 94.40 & 0.46 \\ 
  (1000, 1000) & MCAR & $\mc S$-score (BT-Wald) & 0.18 & 0.054 & 95.15 & 0.21 & 0.06 & 0.119 & 94.40 & 0.46 \\ 
  Large & (40\%) & $\mc S$-score (BT-QT) & 0.18 & 0.054 & 94.55 & 0.21 & 0.06 & 0.119 & 94.35 & 0.46 \\ 
   &  & Pocock & 0.24 & 0.059 & 94.45 & 0.23 & 7.90 & 0.201 & \textbf{69.90} & 0.45 \\ 
   \addlinespace
   &  & $\mc S$-score (IF-Wald) & 0.51 & 0.055 & 94.85 & 0.21 & 0.01 & 0.118 & 94.45 & 0.46 \\ 
   & MAR & $\mc S$-score (BT-Wald) & 0.51 & 0.055 & 94.85 & 0.21 & 0.01 & 0.118 & 94.70 & 0.46 \\ 
   & (20\%) & $\mc S$-score (BT-QT) & 0.51 & 0.055 & 94.00 & 0.21 & 0.01 & 0.118 & 94.30 & 0.46 \\ 
   &  & Pocock & 5.72 & 0.079 & \textbf{81.70} & 0.21 & 6.97 & 0.185 & \textbf{75.80} & 0.45 \\ 
   \addlinespace
   &  & $\mc S$-score (IF-Wald) & 0.76 & 0.052 & 96.00 & 0.21 & 0.00 & 0.117 & 95.10 & 0.46 \\ 
   & MAR & $\mc S$-score (BT-Wald) & 0.76 & 0.052 & 95.95 & 0.21 & 0.00 & 0.117 & 95.10 & 0.46 \\ 
   & (40\%) & $\mc S$-score (BT-QT) & 0.76 & 0.052 & 95.85 & 0.21 & 0.00 & 0.117 & 95.25 & 0.46 \\ 
   &  & Pocock & 12.23 & 0.132 & \textbf{36.95} & 0.20 & 13.77 & 0.305 & \textbf{29.15} & 0.42 \\ 
   \addlinespace
   \bottomrule
    \end{tabular}
    \caption{Simulation results: low (around 20\%) and homogeneous censoring in two treatment groups.}
    \label{tab:simu-smlC-homo}
    \begin{tablenotes}
	\scriptsize
	\item CP\%'s outside the interval [94, 96] are highlighted in bold.
     \end{tablenotes}
\end{threeparttable}
\end{table}

\begin{table}[H]
\begin{threeparttable}
    \centering
    \scriptsize
    \begin{tabular}{cclrrrrrrrrrrrrrrrrrrr}
    \toprule
    & & & \multicolumn{4}{c}{$\WR=1$} & \multicolumn{4}{c}{$\WR=2$}\\
    \makecell[l]{$(n_a, n_b)$} & \makecell[c]{Missing \\ data of $Y_2$} & Method  & ARB\% & RMSE & CP\% & Width & ARB\% & RMSE & CP\% & Width \\
    \cmidrule(lr){1-3}\cmidrule(lr){4-7}\cmidrule(lr){8-11}
   &  & $\mc S$-score (IF-Wald) & 1.90 & 0.199 & 95.10 & 0.77 & 2.42 & 0.424 & 95.25 & 1.63 \\ 
   & No & $\mc S$-score (BT-Wald) & 1.90 & 0.199 & \textbf{96.15} & 0.80 & 2.42 & 0.424 & \textbf{96.40} & 1.74 \\ 
   &  & $\mc S$-score (BT-QT) & 1.90 & 0.199 & 95.00 & 0.79 & 2.42 & 0.424 & 95.25 & 1.71 \\ 
   &  & Pocock & 2.28 & 0.224 & 94.95 & 0.88 & 0.43 & 0.465 & 94.85 & 1.82 \\ 
   \addlinespace
   &  & $\mc S$-score (IF-Wald) & 2.54 & 0.199 & 95.30 & 0.77 & 1.73 & 0.422 & 94.45 & 1.61 \\ 
   & MCAR & $\mc S$-score (BT-Wald) & 2.54 & 0.199 & 95.70 & 0.81 & 1.73 & 0.422 & 95.65 & 1.73 \\ 
   & (20\%) & $\mc S$-score (BT-QT) & 2.54 & 0.199 & 95.20 & 0.80 & 1.73 & 0.422 & 95.00 & 1.70 \\ 
   &  & Pocock & 3.26 & 0.227 & 95.25 & 0.90 & 3.18 & 0.448 & \textbf{93.70} & 1.75 \\ 
   \addlinespace
   &  & $\mc S$-score (IF-Wald) & 1.40 & 0.200 & \textbf{93.90} & 0.77 & 2.50 & 0.428 & 94.75 & 1.63 \\ 
  (100, 100) & MCAR & $\mc S$-score (BT-Wald) & 1.40 & 0.200 & 94.95 & 0.80 & 2.50 & 0.428 & 95.90 & 1.74 \\ 
  Small & (40\%) & $\mc S$-score (BT-QT) & 1.40 & 0.200 & 94.65 & 0.79 & 2.50 & 0.428 & 94.80 & 1.72 \\ 
   &  & Pocock & 2.25 & 0.228 & 95.25 & 0.91 & 4.78 & 0.455 & \textbf{93.15} & 1.73 \\ 
   \addlinespace
   &  & $\mc S$-score (IF-Wald) & 3.13 & 0.204 & 95.10 & 0.78 & 2.24 & 0.425 & \textbf{93.65} & 1.62 \\ 
   & MAR & $\mc S$-score (BT-Wald) & 3.13 & 0.204 & 95.85 & 0.82 & 2.24 & 0.425 & 95.05 & 1.74 \\ 
   & (20\%) & $\mc S$-score (BT-QT) & 3.13 & 0.204 & 94.65 & 0.81 & 2.24 & 0.425 & 94.30 & 1.71 \\ 
   &  & Pocock & 1.06 & 0.219 & 94.25 & 0.87 & 4.42 & 0.451 & \textbf{92.60} & 1.73 \\
   \addlinespace
   &  & $\mc S$-score (IF-Wald) & 2.82 & 0.198 & 95.30 & 0.77 & 2.60 & 0.433 & 94.95 & 1.63 \\ 
   & MAR & $\mc S$-score (BT-Wald) & 2.82 & 0.198 & 96.00 & 0.82 & 2.60 & 0.433 & 95.85 & 1.74 \\ 
   & (40\%) & $\mc S$-score (BT-QT) & 2.82 & 0.198 & 94.90 & 0.81 & 2.60 & 0.433 & 94.50 & 1.72 \\ 
   &  & Pocock & 6.37 & 0.213 & 94.35 & 0.83 & 8.61 & 0.465 & \textbf{90.70} & 1.67 \\
   \addlinespace
   \midrule
   &  & $\mc S$-score (IF-Wald) & 0.27 & 0.061 & 94.15 & 0.24 & 0.14 & 0.131 & 94.15 & 0.50 \\ 
   & No & $\mc S$-score (BT-Wald) & 0.27 & 0.061 & 94.30 & 0.24 & 0.14 & 0.131 & 94.05 & 0.50 \\ 
   &  & $\mc S$-score (BT-QT) & 0.27 & 0.061 & 94.20 & 0.23 & 0.14 & 0.131 & \textbf{93.85} & 0.49 \\ 
   &  & Pocock & 0.33 & 0.068 & 94.55 & 0.26 & 2.70 & 0.152 & \textbf{92.05} & 0.53 \\ 
   \addlinespace
   &  & $\mc S$-score (IF-Wald) & 0.37 & 0.061 & 95.30 & 0.24 & 0.16 & 0.129 & 94.70 & 0.50 \\ 
   & MCAR & $\mc S$-score (BT-Wald) & 0.37 & 0.061 & 95.75 & 0.24 & 0.16 & 0.129 & 94.95 & 0.50 \\ 
   & (20\%) & $\mc S$-score (BT-QT) & 0.37 & 0.061 & 94.70 & 0.23 & 0.16 & 0.129 & 94.85 & 0.49 \\ 
   &  & Pocock & 0.42 & 0.069 & 95.00 & 0.27 & 5.11 & 0.172 & \textbf{87.05} & 0.52 \\ 
   \addlinespace
   &  & $\mc S$-score (IF-Wald) & 0.03 & 0.059 & 94.70 & 0.24 & 0.16 & 0.126 & 94.80 & 0.50 \\ 
  (1000, 1000) & MCAR & $\mc S$-score (BT-Wald) & 0.03 & 0.059 & 94.70 & 0.24 & 0.16 & 0.126 & 95.05 & 0.50 \\ 
  Large & (40\%) & $\mc S$-score (BT-QT) & 0.03 & 0.059 & 94.65 & 0.23 & 0.16 & 0.126 & 94.80 & 0.49 \\ 
   &  & Pocock & 0.05 & 0.069 & 94.80 & 0.27 & 7.57 & 0.204 & \textbf{77.30} & 0.51 \\
   \addlinespace
   &  & $\mc S$-score (IF-Wald) & 0.66 & 0.062 & 95.45 & 0.24 & 0.27 & 0.128 & 94.65 & 0.50 \\ 
   & MAR & $\mc S$-score (BT-Wald) & 0.66 & 0.062 & 95.60 & 0.24 & 0.27 & 0.128 & 94.85 & 0.50 \\ 
   & (20\%) & $\mc S$-score (BT-QT) & 0.66 & 0.062 & 94.70 & 0.24 & 0.27 & 0.128 & 94.80 & 0.49 \\ 
   &  & Pocock & 3.86 & 0.077 & \textbf{90.80} & 0.26 & 6.86 & 0.194 & \textbf{80.10} & 0.51 \\
   \addlinespace
   &  & $\mc S$-score (IF-Wald) & 0.56 & 0.059 & 94.90 & 0.24 & 0.22 & 0.126 & 94.90 & 0.50 \\ 
   & MAR & $\mc S$-score (BT-Wald) & 0.56 & 0.059 & 95.10 & 0.24 & 0.22 & 0.126 & 95.35 & 0.50 \\ 
   & (40\%) & $\mc S$-score (BT-QT) & 0.56 & 0.059 & 94.90 & 0.24 & 0.22 & 0.126 & 94.85 & 0.49 \\ 
   &  & Pocock & 8.66 & 0.107 & \textbf{73.65} & 0.25 & 11.20 & 0.265 & \textbf{55.85} & 0.49 \\
   \addlinespace
   \bottomrule
    \end{tabular}
    \caption{Simulation results: moderate (around 40\%) and homogeneous censoring in two treatment groups.}
    \label{tab:simu-larC-homo}
       \begin{tablenotes}
	\scriptsize
	\item CP\%'s outside the interval [94, 96] are highlighted in bold.
     \end{tablenotes}
\end{threeparttable}
\end{table}

\begin{table}[H]
\begin{threeparttable}
    \centering
    \scriptsize
    \begin{tabular}{cclrrrrrrrrrrrrrrrrrrr}
    \toprule
    & & & \multicolumn{4}{c}{$\WR=1$} & \multicolumn{4}{c}{$\WR=2$}\\
    \makecell[l]{$(n_a, n_b)$} & \makecell[c]{Missing \\ data of $Y_2$} & Method  & ARB\% & RMSE & CP\% & Width & ARB\% & RMSE & CP\% & Width \\
    \cmidrule(lr){1-3}\cmidrule(lr){4-7}\cmidrule(lr){8-11}
   &  & $\mc S$-score (IF-Wald) & 1.81 & 0.179 & 95.05 & 0.69 & 1.91 & 0.383 & 94.85 & 1.51 \\ 
   & No & $\mc S$-score (BT-Wald) & 1.81 & 0.179 & 95.55 & 0.72 & 1.91 & 0.383 & 95.80 & 1.60 \\ 
   &  & $\mc S$-score (BT-QT) & 1.81 & 0.179 & 94.20 & 0.71 & 1.91 & 0.383 & 95.40 & 1.58 \\ 
   &  & Pocock & 4.46 & 0.183 & \textbf{93.80} & 0.70 & 0.36 & 0.392 & 95.40 & 1.59 \\
   \addlinespace
   &  & $\mc S$-score (IF-Wald) & 0.93 & 0.179 & \textbf{93.65} & 0.69 & 1.98 & 0.386 & 95.35 & 1.51 \\ 
   & MCAR & $\mc S$-score (BT-Wald) & 0.93 & 0.179 & 94.50 & 0.72 & 1.98 & 0.386 & \textbf{96.30} & 1.60 \\ 
   & (20\%) & $\mc S$-score (BT-QT) & 0.93 & 0.179 & 94.20 & 0.71 & 1.98 & 0.386 & 95.60 & 1.58 \\ 
   &  & Pocock & 3.13 & 0.189 & \textbf{93.75} & 0.73 & 2.35 & 0.395 & 95.05 & 1.57 \\
   \addlinespace
   &  & $\mc S$-score (IF-Wald) & 1.67 & 0.178 & 95.05 & 0.70 & 2.50 & 0.399 & 95.00 & 1.52 \\ 
  (100, 100) & MCAR & $\mc S$-score (BT-Wald) & 1.67 & 0.178 & 95.15 & 0.72 & 2.50 & 0.399 & 95.65 & 1.61 \\ 
  Small & (40\%) & $\mc S$-score (BT-QT) & 1.67 & 0.178 & 95.00 & 0.71 & 2.50 & 0.399 & 94.65 & 1.59 \\ 
   &  & Pocock & 1.03 & 0.194 & 94.95 & 0.77 & 4.37 & 0.400 & \textbf{93.15} & 1.54 \\
   \addlinespace
   &  & $\mc S$-score (IF-Wald) & 2.09 & 0.182 & 94.60 & 0.70 & 2.76 & 0.398 & 95.05 & 1.53 \\ 
   & MAR & $\mc S$-score (BT-Wald) & 2.09 & 0.182 & 95.10 & 0.72 & 2.76 & 0.398 & \textbf{96.15} & 1.62 \\ 
   & (20\%) & $\mc S$-score (BT-QT) & 2.09 & 0.182 & 94.20 & 0.71 & 2.76 & 0.398 & 94.40 & 1.60 \\ 
   &  & Pocock & 7.46 & 0.192 & \textbf{92.25} & 0.69 & 4.02 & 0.397 & \textbf{93.80} & 1.54 \\
   \addlinespace
   &  & $\mc S$-score (IF-Wald) & 2.70 & 0.188 & 94.90 & 0.70 & 2.75 & 0.392 & 95.45 & 1.53 \\ 
   & MAR & $\mc S$-score (BT-Wald) & 2.70 & 0.188 & 95.50 & 0.73 & 2.75 & 0.392 & \textbf{96.45} & 1.62 \\ 
   & (40\%) & $\mc S$-score (BT-QT) & 2.70 & 0.188 & 94.45 & 0.72 & 2.75 & 0.392 & 94.65 & 1.60 \\ 
   &  & Pocock & 10.40 & 0.202 & \textbf{89.50} & 0.68 & 8.94 & 0.409 & \textbf{90.70} & 1.47 \\
   \addlinespace
   \midrule
   &  & $\mc S$-score (IF-Wald) & 0.25 & 0.056 & 95.30 & 0.21 & 0.03 & 0.118 & 94.60 & 0.46 \\ 
   & No & $\mc S$-score (BT-Wald) & 0.25 & 0.056 & 95.25 & 0.22 & 0.03 & 0.118 & 94.80 & 0.47 \\ 
   &  & $\mc S$-score (BT-QT) & 0.25 & 0.056 & 94.60 & 0.21 & 0.03 & 0.118 & 94.35 & 0.46 \\ 
   &  & Pocock & 6.00 & 0.082 & \textbf{79.35} & 0.21 & 2.35 & 0.131 & \textbf{92.35} & 0.48 \\
   \addlinespace
   &  & $\mc S$-score (IF-Wald) & 0.25 & 0.055 & 95.20 & 0.21 & 0.05 & 0.119 & 95.00 & 0.46 \\ 
   & MCAR & $\mc S$-score (BT-Wald) & 0.25 & 0.055 & 95.35 & 0.22 & 0.05 & 0.119 & 94.85 & 0.47 \\ 
   & (20\%) & $\mc S$-score (BT-QT) & 0.25 & 0.055 & 94.55 & 0.21 & 0.05 & 0.119 & 94.80 & 0.46 \\ 
   &  & Pocock & 3.72 & 0.068 & \textbf{91.15} & 0.22 & 4.58 & 0.152 & \textbf{86.85} & 0.47 \\
   \addlinespace
   &  & $\mc S$-score (IF-Wald) & 0.17 & 0.055 & 94.70 & 0.21 & 0.05 & 0.119 & 94.25 & 0.46 \\ 
  (1000, 1000) & MCAR & $\mc S$-score (BT-Wald) & 0.17 & 0.055 & 94.60 & 0.22 & 0.05 & 0.119 & 94.25 & 0.47 \\ 
  Large & (40\%) & $\mc S$-score (BT-QT) & 0.17 & 0.055 & 94.20 & 0.21 & 0.05 & 0.119 & 94.05 & 0.46 \\ 
   &  & Pocock & 1.04 & 0.061 & 94.10 & 0.23 & 6.94 & 0.186 & \textbf{76.30} & 0.46 \\ 
   \addlinespace
   &  & $\mc S$-score (IF-Wald) & 0.58 & 0.056 & 95.10 & 0.22 & 0.32 & 0.120 & 94.75 & 0.47 \\ 
   & MAR & $\mc S$-score (BT-Wald) & 0.58 & 0.056 & 95.20 & 0.22 & 0.32 & 0.120 & 94.70 & 0.47 \\ 
   & (20\%) & $\mc S$-score (BT-QT) & 0.58 & 0.056 & 94.25 & 0.21 & 0.32 & 0.120 & 94.10 & 0.46 \\ 
   &  & Pocock & 8.86 & 0.104 & \textbf{63.60} & 0.21 & 6.46 & 0.179 & \textbf{78.85} & 0.46 \\ 
   \addlinespace
   &  & $\mc S$-score (IF-Wald) & 0.67 & 0.057 & 94.70 & 0.22 & 0.15 & 0.119 & 95.15 & 0.46 \\ 
   & MAR & $\mc S$-score (BT-Wald) & 0.67 & 0.057 & 94.90 & 0.22 & 0.15 & 0.119 & 94.90 & 0.47 \\ 
   & (40\%) & $\mc S$-score (BT-QT) & 0.67 & 0.057 & 94.50 & 0.22 & 0.15 & 0.119 & 94.35 & 0.46 \\ 
   &  & Pocock & 12.21 & 0.133 & \textbf{39.95} & 0.21 & 11.39 & 0.262 & \textbf{45.65} & 0.44 \\
   \addlinespace
  \bottomrule
  \end{tabular}
    \caption{Simulation results: low (around 20\%) and heterogeneous censoring in both groups.}
    \label{tab:simu-smlC-hete}
     \begin{tablenotes}
	\scriptsize
	\item CP\%'s outside the interval [94, 96] are highlighted in bold.
     \end{tablenotes}
    \end{threeparttable}
\end{table}

\begin{table}[H]
    \begin{threeparttable}
    \centering
    \scriptsize
    \begin{tabular}{cclrrrrrrrrrrrrrrrrrrr}
    \toprule
    & & & \multicolumn{4}{c}{$\WR=1$} & \multicolumn{4}{c}{$\WR=2$}\\
    \makecell[l]{$(n_a, n_b)$} & \makecell[c]{Missing \\ data of $Y_2$} & Method  & ARB\% & RMSE & CP\% & Width & ARB\% & RMSE & CP\% & Width \\
    \cmidrule(lr){1-3}\cmidrule(lr){4-7}\cmidrule(lr){8-11}
   &  & $\mc S$-score (IF-Wald) & 2.07 & 0.205 & 94.60 & 0.79 & 2.80 & 0.441 & 94.45 & 1.65 \\ 
   & No & $\mc S$-score (BT-Wald) & 2.07 & 0.205 & 95.70 & 0.83 & 2.80 & 0.441 & 95.20 & 1.77 \\ 
   &  & $\mc S$-score (BT-QT) & 2.07 & 0.205 & 94.35 & 0.82 & 2.80 & 0.441 & \textbf{93.75} & 1.74 \\ 
   &  & Pocock & 5.55 & 0.218 & \textbf{93.40} & 0.84 & 2.69 & 0.468 & \textbf{92.60} & 1.80 \\ 
   \addlinespace
   &  & $\mc S$-score (IF-Wald) & 2.68 & 0.206 & 95.30 & 0.79 & 2.76 & 0.438 & 94.45 & 1.65 \\ 
   & MCAR & $\mc S$-score (BT-Wald) & 2.68 & 0.206 & \textbf{96.45} & 0.84 & 2.76 & 0.438 & 95.75 & 1.77 \\ 
   & (20\%) & $\mc S$-score (BT-QT) & 2.68 & 0.206 & 95.45 & 0.83 & 2.76 & 0.438 & 94.60 & 1.74 \\ 
   &  & Pocock & 2.95 & 0.219 & 95.25 & 0.87 & 3.55 & 0.469 & \textbf{93.25} & 1.79 \\ 
   \addlinespace
   &  & $\mc S$-score (IF-Wald) & 1.96 & 0.208 & 94.25 & 0.79 & 1.83 & 0.426 & \textbf{93.60} & 1.63 \\ 
  (100, 100) & MCAR & $\mc S$-score (BT-Wald) & 1.96 & 0.208 & \textbf{96.15} & 0.84 & 1.83 & 0.426 & 94.75 & 1.75 \\ 
  Small & (40\%) & $\mc S$-score (BT-QT) & 1.96 & 0.208 & 94.80 & 0.83 & 1.83 & 0.426 & 94.15 & 1.72 \\ 
   &  & Pocock & 0.84 & 0.228 & 94.40 & 0.90 & 5.50 & 0.458 & \textbf{92.70} & 1.75 \\ 
   \addlinespace
   &  & $\mc S$-score (IF-Wald) & 2.72 & 0.208 & 94.85 & 0.79 & 2.27 & 0.446 & 94.35 & 1.64 \\ 
   & MAR & $\mc S$-score (BT-Wald) & 2.72 & 0.208 & 95.75 & 0.85 & 2.27 & 0.446 & 95.75 & 1.76 \\ 
   & (20\%) & $\mc S$-score (BT-QT) & 2.72 & 0.208 & 95.05 & 0.84 & 2.27 & 0.446 & 94.35 & 1.73 \\ 
   &  & Pocock & 6.57 & 0.219 & \textbf{93.65} & 0.83 & 5.34 & 0.477 & \textbf{92.50} & 1.75 \\ 
   \addlinespace
   &  & $\mc S$-score (IF-Wald) & 4.08 & 0.216 & 94.45 & 0.81 & 2.99 & 0.445 & 94.50 & 1.65 \\ 
   & MAR & $\mc S$-score (BT-Wald) & 4.08 & 0.216 & 95.70 & 0.87 & 2.99 & 0.445 & 95.75 & 1.77 \\ 
   & (40\%) & $\mc S$-score (BT-QT) & 4.08 & 0.216 & 94.30 & 0.86 & 2.99 & 0.445 & \textbf{93.90} & 1.75 \\ 
   &  & Pocock & 8.29 & 0.224 & \textbf{92.10} & 0.83 & 7.08 & 0.467 & \textbf{91.35} & 1.72 \\ 
   \addlinespace
   \midrule
   &  & $\mc S$-score (IF-Wald) & 0.14 & 0.060 & 95.45 & 0.24 & 0.13 & 0.125 & 94.90 & 0.50 \\ 
   & No & $\mc S$-score (BT-Wald) & 0.14 & 0.060 & 95.45 & 0.24 & 0.13 & 0.125 & 95.25 & 0.50 \\ 
   &  & $\mc S$-score (BT-QT) & 0.14 & 0.060 & 95.65 & 0.24 & 0.13 & 0.125 & 95.35 & 0.50 \\ 
   &  & Pocock & 7.95 & 0.102 & \textbf{76.35} & 0.25 & 5.92 & 0.180 & \textbf{84.50} & 0.53 \\ 
   \addlinespace
   &  & $\mc S$-score (IF-Wald) & 0.22 & 0.062 & 95.10 & 0.24 & 0.30 & 0.127 & 95.25 & 0.50 \\ 
   & MCAR & $\mc S$-score (BT-Wald) & 0.22 & 0.062 & 95.15 & 0.24 & 0.30 & 0.127 & 95.35 & 0.50 \\ 
   & (20\%) & $\mc S$-score (BT-QT) & 0.22 & 0.062 & 94.80 & 0.24 & 0.30 & 0.127 & 94.35 & 0.50 \\ 
   &  & Pocock & 5.41 & 0.085 & \textbf{86.80} & 0.26 & 6.57 & 0.189 & \textbf{82.00} & 0.52 \\ 
   \addlinespace
   &  & $\mc S$-score (IF-Wald) & 0.27 & 0.061 & 95.40 & 0.24 & 0.30 & 0.128 & 95.20 & 0.50 \\ 
  (1000, 1000) & MCAR & $\mc S$-score (BT-Wald) & 0.27 & 0.061 & 95.25 & 0.24 & 0.30 & 0.128 & 95.40 & 0.50 \\ 
  Large & (40\%) & $\mc S$-score (BT-QT) & 0.27 & 0.061 & 94.85 & 0.24 & 0.30 & 0.128 & 94.40 & 0.50 \\ 
   &  & Pocock & 2.57 & 0.073 & \textbf{92.80} & 0.27 & 7.53 & 0.205 & \textbf{77.70} & 0.52 \\ 
   \addlinespace
   &  & $\mc S$-score (IF-Wald) & 0.52 & 0.062 & 95.35 & 0.24 & 0.32 & 0.127 & 95.30 & 0.50 \\ 
   & MAR & $\mc S$-score (BT-Wald) & 0.52 & 0.062 & 95.45 & 0.24 & 0.32 & 0.127 & 95.35 & 0.50 \\ 
   & (20\%) & $\mc S$-score (BT-QT) & 0.52 & 0.062 & 94.90 & 0.24 & 0.32 & 0.127 & 94.55 & 0.50 \\ 
   &  & Pocock & 8.86 & 0.109 & \textbf{72.20} & 0.25 & 7.72 & 0.206 & \textbf{77.45} & 0.52 \\ 
   \addlinespace
   &  & $\mc S$-score (IF-Wald) & 0.75 & 0.062 & 95.50 & 0.24 & 0.38 & 0.128 & 95.30 & 0.50 \\ 
   & MAR & $\mc S$-score (BT-Wald) & 0.75 & 0.062 & 95.55 & 0.25 & 0.38 & 0.128 & 95.30 & 0.50 \\ 
   & (40\%) & $\mc S$-score (BT-QT) & 0.75 & 0.062 & 95.00 & 0.24 & 0.38 & 0.128 & 94.45 & 0.50 \\ 
   &  & Pocock & 9.89 & 0.117 & \textbf{67.20} & 0.25 & 9.76 & 0.241 & \textbf{65.60} & 0.51 \\ 
   \addlinespace
  \bottomrule
  \end{tabular}
    \caption{Simulation results: moderate (around 40\%) and heterogeneous censoring in both groups.}
    \label{tab:simu-larC-hete}
        \begin{tablenotes}
	\scriptsize
	\item CP\%'s outside the interval [94, 96] are highlighted in bold.
     \end{tablenotes}
    \end{threeparttable}
\end{table}

\subsection{Additional experiment on coverage probability under small sample size}

\begin{table}[H]
    \begin{threeparttable}
    \centering
    \scriptsize
    \begin{tabular}{llccccccccccccccc}
    \toprule
    & & \multicolumn{3}{c}{$\WR=1$} & \multicolumn{3}{c}{$\WR=2$}\\
    \makecell[c]{Censoring of $Y_1$} & \makecell[c]{Missingness of $Y_2$}  & IF-Wald & BT-Wald & BT-QT & IF-Wald & BT-Wald & BT-QT \\
    \cmidrule(lr){1-2}\cmidrule(lr){3-5}\cmidrule(lr){6-8}
                                & No & \textbf{92.95} & 94.40 & 94.20 & \textbf{92.80} & 94.80 & \textbf{93.85} \\
   No censored $Y_1$ & MCAR (40\%) & \textbf{92.10} & 94.50 & 94.10 & \textbf{93.55} & 95.50 & \textbf{93.55} \\
                    & MAR (40\%) &\textbf{93.85} & 95.20 & 94.10 & \textbf{93.25} & 95.60 & 95.10  \\
   \midrule
                                & No & \textbf{93.60} & 94.70 & 95.40 & \textbf{91.90} & 94.60 & 94.25 \\
   40\% censored $Y_1$, & MCAR (40\%) & \textbf{92.75} & 95.50 & 94.80 & \textbf{91.80} & 95.60 & 94.55 \\
   homogeneous across groups & MAR (40\%) & 94.40 & \textbf{96.40} & 95.70 & \textbf{92.85} & 95.70 & 94.90 \\
   \midrule
                                    & No & \textbf{92.05} & 95.35 & 95.20 & \textbf{92.80} & 94.75 & 95.10 \\
   40\% censored $Y_1$, & MCAR (40\%) & \textbf{93.75} & \textbf{96.40} & 94.80 & \textbf{93.55} & 95.10 & 94.10 \\
   heterogeneous across groups & MAR (40\%) & \textbf{93.40} & 96.00 & 95.20 & \textbf{92.50} & 95.80 & 95.10 \\
    \bottomrule
    \end{tabular}
      \caption{Coverage probabilities (CP\%) of IF-based and bootstrap-based confidence interval (CI) estimators for the $\mc S$-score across the simulation scenarios in the main text, where $n_a=n_b=25$. CP\% values outside [94, 96] range are marked in bold. }\label{tab:cp-smallsample}
\end{threeparttable}
\end{table}

\section{Additional Data Analysis Results}\label{app:data}

\subsection{Sensitivity analysis of varying MAR models in the HEART-FID study}\label{subapp:sens-fid}

\begin{table}[H]
\begin{threeparttable}
    \centering
    \footnotesize
    \begin{tabular}{llccc}
    \toprule
   Missingness model &  Method & Estimate & 95\% CI & CI Width \\ 
      \midrule
 \multirow{3}{*}{Logistic regression} 
 & Covariate-adjusted $\mc S$-score (IF-Wald) & 1.02 & (0.94, 1.11) & 0.18 \\ 
 & Covariate-adjusted $\mc S$-score (BT-Wald) & 1.02 & (0.93, 1.12) & 0.18 \\ 
 & Covariate-adjusted $\mc S$-score (BT-QT) & 1.02 & (0.94, 1.12) & 0.18 \\ 
  \midrule
  \multirow{3}{*}{Logistic regression (interaction)} 
 & Covariate-adjusted $\mc S$-score (IF-Wald) & 1.02 & (0.94, 1.11) & 0.18 \\ 
 & Covariate-adjusted $\mc S$-score (BT-Wald) & 1.02 & (0.94, 1.11) & 0.18 \\ 
 & Covariate-adjusted $\mc S$-score (BT-QT) & 1.02 & (0.94, 1.11) & 0.17 \\ 
  \midrule
  \multirow{3}{*}{Random forest} 
 & Covariate-adjusted $\mc S$-score (IF-Wald) & 1.02 & (0.93, 1.11) & 0.18 \\ 
 & Covariate-adjusted $\mc S$-score (BT-Wald) & 1.02 & (0.93, 1.10) & 0.18 \\ 
 & Covariate-adjusted $\mc S$-score (BT-QT) & 1.02 & (0.93, 1.11) & 0.18 \\  
 \midrule
 \multirow{3}{*}{Ensemble learning}
 & Covariate-adjusted $\mc S$-score (IF-Wald) & 1.02 & (0.93, 1.11) & 0.18 \\ 
 & Covariate-adjusted $\mc S$-score (BT-Wald) & 1.02 & (0.93, 1.11) & 0.17 \\ 
 & Covariate-adjusted $\mc S$-score (BT-QT) & 1.02 & (0.94, 1.11) & 0.17 \\ 
       \bottomrule
    \end{tabular}
    \caption{HEART-FID trial: Additional sensitivity analyses of win ratio estimates for time to death and the 6-minute walk distance at month 12 under under different missingness models for the second endpoint.  }
    \label{tab:dataI-sens}
     \begin{tablenotes}
	\scriptsize
	\item CI: confidence interval.
    \item Logistic regression (interaction): logistic regression model including all pairwise interaction terms among covariates.
    \item Ensemble learning: an ensemble model including logistic regression, logistic regression (interaction) and random forest by \texttt{SuperLearner} R package. 
     \end{tablenotes}
\end{threeparttable}
\end{table}

\subsection{Sensitivity analysis of varying MAR models in the ISCHEMIA study}\label{subapp:sens-isch}
\begin{table}[H]
\begin{threeparttable}
    \centering
    \footnotesize
    \begin{tabular}{llccc}
    \toprule
   Missingness model &  Method & Estimate & 95\% CI & CI Width \\ 
      \midrule
 \multirow{3}{*}{Logistic regression} 
 & Covariate-adjusted $\mc S$-score (IF-Wald) & 1.27 & (1.05, 1.49) & 0.44 \\ 
 &  Covariate-adjusted $\mc S$-score (BT-Wald) & 1.27 & (1.05, 1.49) & 0.44 \\ 
 &  Covariate-adjusted $\mc S$-score (BT-QT) & 1.27 & (1.08, 1.50) & 0.42 \\
  \midrule
  \multirow{3}{*}{Logistic regression (interaction)} 
 &  Covariate-adjusted $\mc S$-score (IF-Wald) & 1.28 & (1.05, 1.50) & 0.45 \\ 
 &  Covariate-adjusted $\mc S$-score (BT-Wald) & 1.28 & (1.05, 1.40) & 0.45 \\ 
 &  Covariate-adjusted $\mc S$-score (BT-QT) & 1.28 & (1.08, 1.53) & 0.45 \\
  \midrule
  \multirow{3}{*}{Random forest} 
 &  Covariate-adjusted $\mc S$-score (IF-Wald) & 1.27 & (1.05, 1.49) & 0.44 \\ 
 &  Covariate-adjusted $\mc S$-score (BT-Wald) & 1.27 & (1.06, 1.48) & 0.42 \\ 
 &  Covariate-adjusted $\mc S$-score (BT-QT) & 1.27 & (1.09, 1.51) & 0.43 \\
  \midrule
 \multirow{3}{*}{Ensemble learning}
 &  Covariate-adjusted $\mc S$-score (IF-Wald) & 1.27 & (1.05, 1.49) & 0.44 \\ 
 &  Covariate-adjusted $\mc S$-score (BT-Wald) & 1.27 & (1.06, 1.48) & 0.43 \\ 
 &  Covariate-adjusted $\mc S$-score (BT-QT) & 1.27 & (1.08, 1.50) & 0.42 \\
       \bottomrule
    \end{tabular}
    \caption{ISCHEMIA trial: Additional sensitivity analyses of win ratio estimates for time to death and the SAQ7-AF score at month 48 under different missingness models for the second endpoint. }
    \label{tab:dataII-sens}
     \begin{tablenotes}
	\scriptsize
	\item CI: confidence interval.
    \item Logistic regression (interaction): logistic regression model including all pairwise interaction terms among covariates.
    \item Ensemble learning: an ensemble model including logistic regression, logistic regression (interaction) and random forest by \texttt{SuperLearner} R package. 
     \end{tablenotes}
\end{threeparttable}
\end{table}

\end{document}